\documentclass[aps,prb,twocolumn,superscriptaddress,english]{revtex4-2}
\usepackage[utf8]{inputenc}

\usepackage{amsmath}
\usepackage{amssymb}
\usepackage{booktabs}
\usepackage{dirtytalk}
\usepackage{graphicx}
\usepackage{xcolor}
\usepackage{lipsum}
\usepackage{lineno}

\usepackage[version=4]{mhchem}
\usepackage{chemformula}
\usepackage{isotope}

\usepackage{amssymb}
\usepackage{amsmath}
\usepackage{mathtools}
\usepackage{physics}

\usepackage{hyperref}

\makeatletter
\def\@fnsymbol#1{\ensuremath{\ifcase#1\or \dagger\or *\or \ddagger\or
   \mathsection\or \mathparagraph\or \|\or **\or \dagger\dagger
   \or \ddagger\ddagger \else\@ctrerr\fi}}
\makeatother

\usepackage{xr-hyper}
\newcommand*{\addFileDependency}[1]{
\typeout{(#1)}
\@addtofilelist{#1}
\IfFileExists{#1}{}{\typeout{No file #1.}}
}\makeatother

\graphicspath{{plots/}}

\begin{document}

\author{Alessio Cucciari}\email{alessio.cucciari@uniroma1.it}
\affiliation{Dipartimento di Fisica, Sapienza - Universit\`a di Roma, 00185 Rome, Italy}
\author{Lilia Boeri} \email{lilia.boeri@uniroma1.it}
\affiliation{Dipartimento di Fisica, Sapienza - Universit\`a di Roma, 00185 Rome, Italy}

\title{An \textit{ab initio} answer to long-debated questions about superconducting Nb$_3$Sn}

\date{\today}
\begin{abstract}
We present the first fully \textit{ab initio} microscopic description of cubic and tetragonal Nb$_3$Sn. We compute the anharmonic free energy surface, phonon spectra, and solve the full-bandwidth anisotropic Migdal–Eliashberg equations for the superconducting gap of the two phases.
Our results show that anharmonic effects are crucial to stabilize both the cubic and tetragonal structures, yielding phonon spectra in excellent agreement with neutron scattering data. We find that the martensitic transition is weakly first-order and that the superconducting gap is strongly anisotropic yet fully-open, with contributions from both longitudinal and transverse Nb $d$-orbitals, revealing an unexpected three-dimensional pairing mechanism. We also find that the experimentally observed reduction of the upper critical field $H_{c2}$ across the transition is explained by a combination of overall weaker electron–phonon coupling and a redistribution of Fermi velocities, which shifts parts of the Fermi surface to longer coherence lengths and limits $H_{c2}$. Based on these insights, we propose that Sn-site doping could enhance transverse-state coupling and gap isotropy, potentially improving both $T_c$ and $H_{c2}$, while Nb-site doping reinforce $H_{c2}$ at the cost of lowering $T_c$.
\end{abstract}

\maketitle



\section{Introduction}

Whether or not the global race toward room-temperature superconductivity succeeds, many technologies will still rely on materials that operate under strong magnetic fields, rather than merely at high temperatures. These include magnetic resonance imaging (MRI), particle accelerators and fusion reactors, which demand superconductors with high upper critical fields $H_{c2}$, large critical current densities $J_c$ and scalable manufacturing.
\\
\indent
High-$T_c$ cuprates like rare-earth barium copper oxides (REBCO) offer impressive performance, with $H_{c2}$ exceeding 100 T \cite{Larbalestier_review_nature_2001}, but strong anisotropy and fabrication costs have limited their widespread use. For this reason, low-$T_c$ conventional superconductors like NbTi and Nb$_3$Sn remain the most practical choices, even decades after their discovery. Among them, NbTi is more widely used but already operates near its intrinsic limits \cite{Godeke_review_IOP_2006}. Meanwhile, Nb$_3$Sn is the leading candidate for next-generation magnets thanks to its higher $T_c$ (18.3 K) and $H_{c2}$ (up to 29 T) \cite{Yao_review_scmarket_2021}. Despite these advantages, Nb$_3$Sn remained less studied than NbTi and still offers potential for further optimization.
\begin{figure}[b]
	\includegraphics[width=0.8\columnwidth]{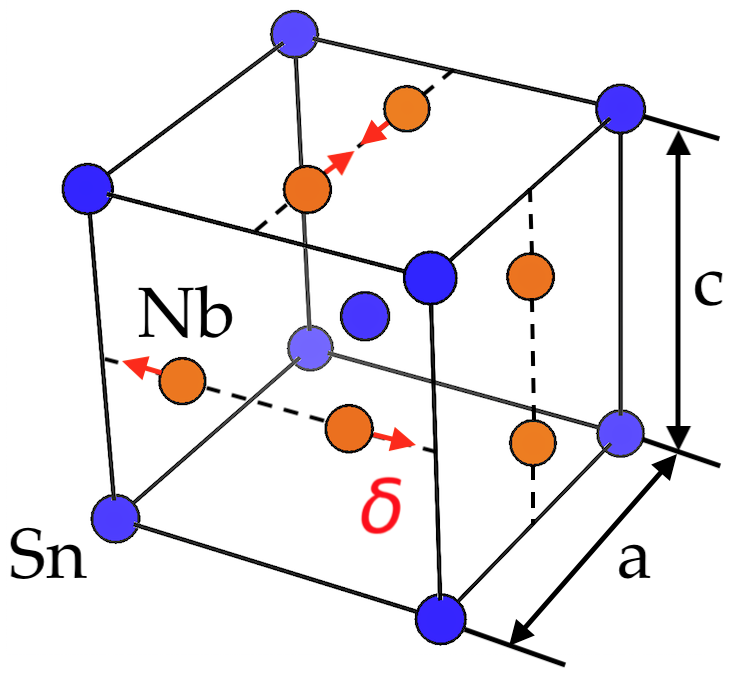}
	\caption{\footnotesize{Crystal structure of A15 Nb$_3$Sn. Nb (orange) and Sn (blue) atoms occupy 6$c$ and 2$a$ Wyckoff positions, respectively.}}
    \label{fig:lattice}
\end{figure}
\\
\indent
Nb$_3$Sn belongs to the A15 family of superconductors \cite{Stewart_review_a15_2015}, a class of intermetallic compounds with general formula A$_3$B and preferred cubic symmetry (space group $Pm\overline{3}n$). These compounds feature a body-centered sublattice of B atoms (e.g., Sn) and three orthogonal chains of A atoms (e.g., Nb) along the cubic axes, as illustrated in Fig. \ref{fig:lattice}. Many A15 superconductors undergo a martensitic transformation \cite{Dew_review_a15_1975}: upon cooling below a temperature $T_m > T_c$, nearly stoichiometric samples undergo a diffusionless structural transition to a tetragonal phase (space group $P4_2/mmc$) \cite{Shirane_nb3sn_delta_PRB_1971}. This transformation is driven by a zone-center optical phonon mode of symmetry $\Gamma_{12}^+$, which induces a dimerization $\delta$ of the chains and couples linearly to a tetragonal strain $\varepsilon = c/a - 1$~\cite{Bhatt_peierls_a15_PRB_1976}. As shown in Fig.~\ref{fig:lattice}, the Nb atoms shift alternately along the chain directions (red arrows), and the lattice distorts along one axis. The transition occurs around $T_m \simeq 43$ K in Nb$_3$Sn, with measured $\varepsilon = -0.006$ and $\delta = -0.003$ \cite{Shirane_nb3sn_delta_PRB_1971, Fuji_nb3sn_delta_PRB_1982}. In tetragonal samples, $T_c$ is reduced by $\sim$ 1 K compared to cubic samples but, more importantly, $H_{c2}$ drops from its optimal value of 29 T to about 21 T \cite{Testardi_review_RMP_1975,Godeke_review_IOP_2006}, suppressing practical performance.
\\
\indent
For this reason, optimization of martensitic Nb$_3$Sn has relied on empirical strategies to prevent the transition. For instance, introducing impurities in small concentrations (such as Ti, Ta or Hf) is know to stabilize the cubic structure and also boost $H_{c2}$, even if it slightly reduces $T_c$ \cite{Tachinawa_nb3sn_ti_1982,Flukiger_nb3sn_wires_2008,Heald_nb3sn_tita_2018,Tarantini_hf_nb3sn_2021}. Instead, theory-guided optimization has been challenging. In fact, \textit{ab initio} Migdal–Eliashberg (ME) calculations, successful for many conventional superconductors \cite{Sanna_PRL_2020_SCDFT_bench}, cannot reproduce the critical temperature of compounds like NbTi and Nb$_3$Sn, because the underlying assumptions -- dynamically stable structure, perfectly stoichiometric crystals -- break down \cite{Cucciari_nbti_PRB_2024, Sadigh_nb3sn_PRB_1998, Klein_double_well_nb3sn_2001}. Some studies circumvented this issue by artificially stabilizing the structure through increased electronic smearing \cite{Tutuncu_nb3sn_phonons_PRB_2006, DeMarzi_nb3sn_strain_IOP_2013, Zhang_nb3sn_pressure_2015, Gala_nb3sn_alloy_2016, Yang_nb3sn_freeenergy_2023,Wu_nb3sn_cdw_2023, Chen_nb3sn_pressure_2025}, but this is physically unjustified. A proper treatment requires going beyond the harmonic approximation and including anharmonic effects. Such calculations were computationally unfeasible until recently, but thanks to the combination of the Stochastic Self-Consistent Harmonic Approximation (SSCHA) with Machine-Learning Interatomic Potentials (MLIPs) \cite{Monacelli_JPCM_2021_SSCHA, Lucrezi_FBW_2024,Ferreira_alloy_2024}, it is now possible to include these effects microscopically, as we have demonstrated for NbTi \cite{Cucciari_nbti_PRB_2024}.
\\
\indent
In this work, we perform state-of-the-art \textit{ab initio} calculations to provide the first, fully microscopic description of the superconducting state of Nb$_3$Sn. To do so, we needed to address several open questions that are critical for guiding material optimization. 
In particular, we address three unresolved issues. First, the thermodynamic nature of the martensitic transition: Anderson and Blount \cite{Anderson_martensitic_PRL_1965} showed that a cubic-to-tetragonal transformation should be first order in the absence of any change in internal symmetry other than mere strain. However, in V$_3$Si the transition is second-order \cite{Batterman_v3si_PRL_1964, Vieland_nb3sn_v3si_tm_1969}, while for Nb$_3$Sn results are unclear \cite{Mailfert_nb3sn_tm_PLA_1967, Vieland_nb3sn_v3si_tm_1969, Escudero_nb3sn_firstorder_2009,Acosta_nb3sn_nbnbchains_2011}. Second, the anisotropy of the superconducting gap: early specific heat and point-contact spectroscopy studies suggested a two-gap scenario \cite{Guritanu_Nb3sn_gap_PRB_2004, Marz_nb3sn_twogap_prb_2010}, while more recent measurements hint at a single gap \cite{Escudero_nb3sn_firstorder_2009,Jo_nb3sn_gap_2014}. A detailed understanding of the gap anisotropy is crucial for wire design, as it governs vortex formation and pinning, with direct impact on both $H_{c2}$ and $J_c$. 
Third, the role of the phase transition in suppressing $H_{c2}$: recent reports challenge the established idea that $H_{c2}$ drops due to the martensitic transition to the tetragonal phase  \cite{Zhou_nb3sn_tetragonal_hc2_2011}, and hence the question of its microscopic origin remains open.
\\
\indent
The paper is organized as follows. In Sec. II, we discuss the nature of the martensitic transition describing the profile of the Born-Oppenheimer surface and its anharmonic generalization. In Sec. III, we examine the electronic structures of the cubic and tetragonal phases. In Sec. IV we report the anharmonic phonon spectra computed via the SSCHA-MLIP method, while Sec. V discusses the superconducting properties obtained from the solution of the anisotropic Migdal–Eliashberg equations and analyses possible optimization strategies. A summary of our key findings is given in Sec. VI. Technical details of the calculations are provided in the Appendix. 

\section{Martensitic transition}

\begin{figure}[t]
	\includegraphics[width=0.9\columnwidth]{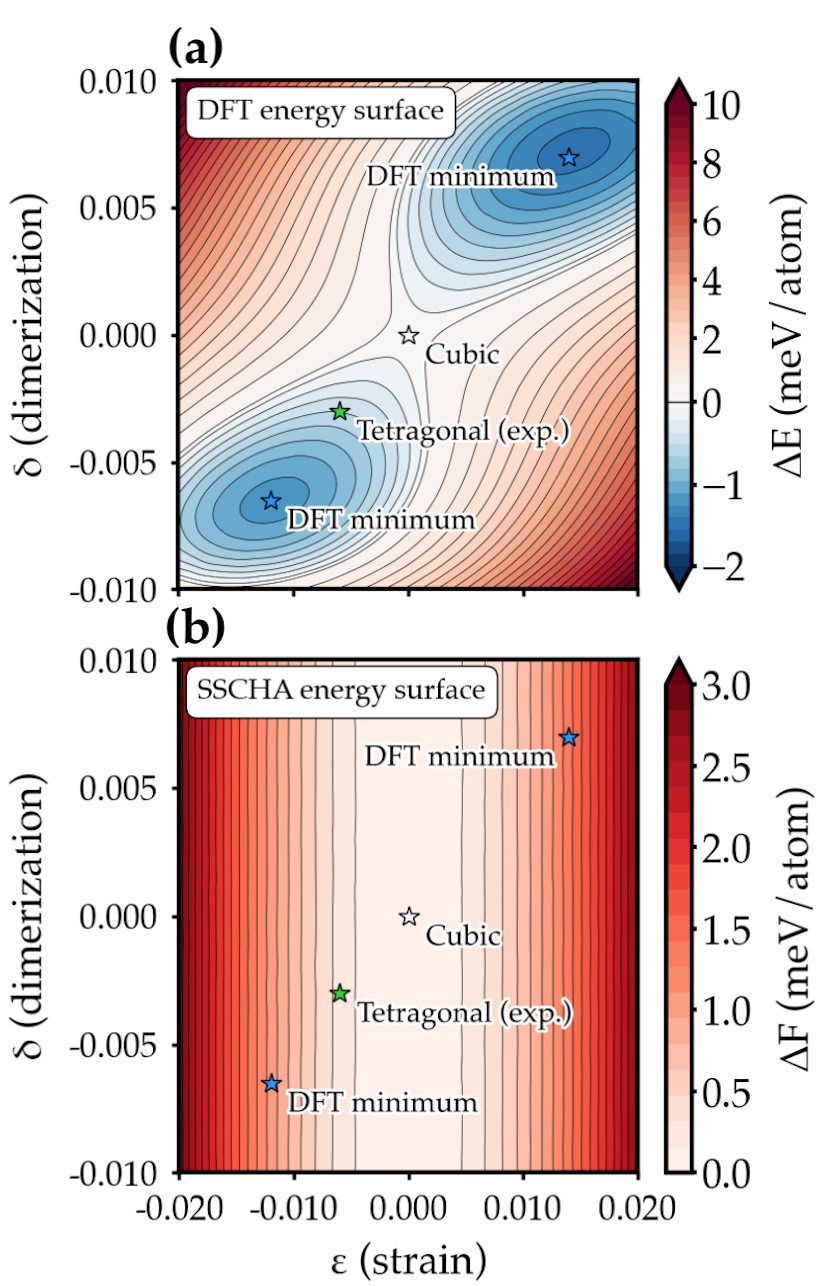}
	\caption{\footnotesize{\textbf{(a)} Contour plot of the fixed-volume Born–Oppenheimer (BO) Potential Energy Surface $\Delta E(\varepsilon, \delta)$ computed with DFT. The energy is given relative to the cubic configuration $(\varepsilon=0,\delta=0)$, which corresponds to a saddle point (white star). Two locally stable minima (blue stars) are located at $(-0.012, -0.006)$ and $(0.014, 0.007)$, while the experimentally observed distortion is indicated by a green star. \textbf{(b)} Same as panel (a), but including zero-point energy corrections through SSCHA. $\Delta F$ denotes the resulting anharmonic free energy difference relative to the cubic configuration, which serves again as a reference.}}
	\label{fig:BO}
\end{figure}

\begin{table*}[t]
\centering
\renewcommand{\arraystretch}{1.5}
\setlength{\tabcolsep}{6pt}
\begin{tabular}{lccccccc}
\hline
\hline
Phase & Space Group & Lattice parameters (\AA) & Atomic positions (Wyckoff) & $x$ & $y$ & $z$ \\
\hline
Cubic     & $Pm\overline{3}n$ (223) & $a = b = c = 5.264$        & Sn ($2a$)  & 0.0000 & 0.0000 & 0.0000 \\
           &                         &                            & Nb ($6c$)  & 0.0000 & 0.5000 & 0.2500 \\
           \hline
Tetragonal & $P4_2/mmc$ (131)        & $a = b = 5.276,\; c = 5.241$ & Nb ($2c$) & 0.0000 & 0.5000 & 0.0000 \\
           &                         &                            & Nb ($2f$) & 0.5000 & 0.5000 & 0.2500 \\
           &                         &                            & Sn ($4j$)  & 0.2468 & 0.0000 & 0.0000 \\
\hline
\hline
\end{tabular}
\caption{Crystallographic data for cubic and tetragonal (experimental) Nb$_3$Sn structures. Lattice parameters for the cubic phase are obtained from SSCHA relaxation at zero pressure. Lattice parameters for the tetragonal phase have been chosen to conserve the same unit cell volume. The table reports the structure phase, space group, lattice parameters (in \AA), atomic site with Wyckoff position in parentheses, and fractional atomic coordinates.}
\label{tab:cryst_struct}
\end{table*}

In order to investigate the origin of the martensitic transition in Nb$_3$Sn, we computed its Born–Oppenheimer (BO) potential energy surface at fixed volume since the real transition only involves minor volume changes that can be safely neglected. We employed Density Functional Theory (DFT) with the PBEsol exchange-correlation functional \cite{Perdew_pbesol_PRL_2008}, which is known to be more accurate than standard PBE. Further computational details are given in the Appendix. Results are shown in Fig.~\ref{fig:BO} (a), where the ground state energy of the system is plotted as a function of two structural order parameters: the tetragonal strain $\varepsilon = c/a - 1$ (horizontal axis) and the Nb-chain dimerization $\delta$ (vertical axis), which corresponds to the eigenvector of the $\Gamma_{12}^+$ optical phonon, as defined in Fig. \ref{fig:lattice}. The color scale indicates the total energy per atom relative to the cubic configuration $(\varepsilon=0,\delta=0)$, with red and blue indicating higher and lower energies, respectively, and thus representing the depth of the energy landscape. Fig.~\ref{fig:BO} (a) reveals a double-well profile, with two slightly asymmetric minima (blue stars) separated by a shallow energy barrier of less than 2 meV/atom. A saddle point (white star) occurs precisely at (0,0), i.e. the cubic phase. This result is in excellent agreement with earlier calculations and confirms that the cubic phase of Nb$_3$Sn is dynamically unstable at the harmonic level \cite{Sadigh_nb3sn_PRB_1998, Klein_double_well_nb3sn_2001}: the unstable $\Gamma_{12}^+$ phonon induces a dimerization $\delta$ of the Nb-chains that makes the system collapse into one of the tetragonal minima. This dynamical instability can be artificially suppressed by increasing the electronic smearing, as shown in Fig. S1 in the Supplemental Material (SM) \cite{suppmat}.
\\
\indent
Based on Fig. \ref{fig:BO} (a), DFT predicts the martensitic transition to be a second-order Peierls-like transition, as proposed by early semi-phenomenological models \cite{Bhatt_peierls_a15_PRB_1976}. DFT also overestimates the equilibrium values of $\varepsilon$ and $\delta$ by approximately a factor of two compared to experiments. However, the energy difference between the theoretical and experimental minima is small, about 0.5 meV/atom, i.e. within the typical accuracy of DFT calculations. We will show in Sec. IV that the tetragonal phase is also dynamically unstable within the harmonic approximation, although its instability is associated with different phonon eigenvectors.
\\
\indent
Since phonons are unstable at the harmonic level, but the crystals are stable experimentally, we employed the Stochastic Self-Consistent Harmonic Approximation (SSCHA) to assess how anharmonicity would restore the correct energy landscape. At $T = 0$~K, anharmonic effects contribute through quantum zero-point fluctuations. The resulting landscape differs qualitatively from the DFT one -- see Fig. \ref{fig:BO} (b): the cubic configuration becomes the global minimum of a broad and shallow well, of depth $\sim$ 3 meV/atom. When tetragonal structures are relaxed within SSCHA, they always relax back to the cubic structure. Thus, within the accuracy of calculations with quantum lattice SSCHA corrections, the cubic structure remains the most stable at all temperatures, even below the experimental transition temperature $T_m$. \\
\indent
Overall, our theoretical result differs from experimental observations, which report a clear structural transition to a tetragonal phase below $T_m$. However, the tetragonal transition is not observed in all samples: it only appears in nearly perfectly stoichiometric samples \cite{Godeke_review_IOP_2006}. Even small deviations from stoichiometry are sufficient to suppress the transition, stabilizing the cubic phase below $T_m$ \cite{Schicktanz_nb3sn_notrans_PRB_1980}. Moreover, within SSCHA all tetragonal configurations ($\varepsilon \neq 0$) exhibit a finite internal stress. In the experimental configuration $(\varepsilon_{exp} = -0.006,\; \delta_{exp} = -0.003)$, the average internal stress is about 0.2 GPa, in good agreement with experiments \cite{Jin_nb3sn_residual_stress_2017}. This suggests that internal pressure, arising for instance by grain boundaries, could help stabilize the tetragonal phase in real samples, even though the cubic structure is energetically preferred. Thus, synthesis methods that minimize stress -- like slow annealing or epitaxial growth -- could prevent the transition. Recent attempts at using epitaxial growth for radio-frequency cavity applications have shown promise \cite{Lee_nb3sn_epitaxial_2019}, but achieving high Sn content remains challenging due to the limited diffusion of Sn without a Cu matrix.
\\
\indent
Since within SSCHA we did not find a minimum for the tetragonal structure \cite{Hybrid_functional_footnote}, in the rest of this work we adopt the experimental tetragonal structure as a reference for comparison with the cubic phase. Hereafter, we will refer to the latter as \textit{the cubic phase} and to the former as \textit{the tetragonal phase}. Crystallographic data are reported in Table \ref{tab:cryst_struct}.

\section{Electronic structure}

\begin{figure}[t]
	\includegraphics[width=1.03\columnwidth]{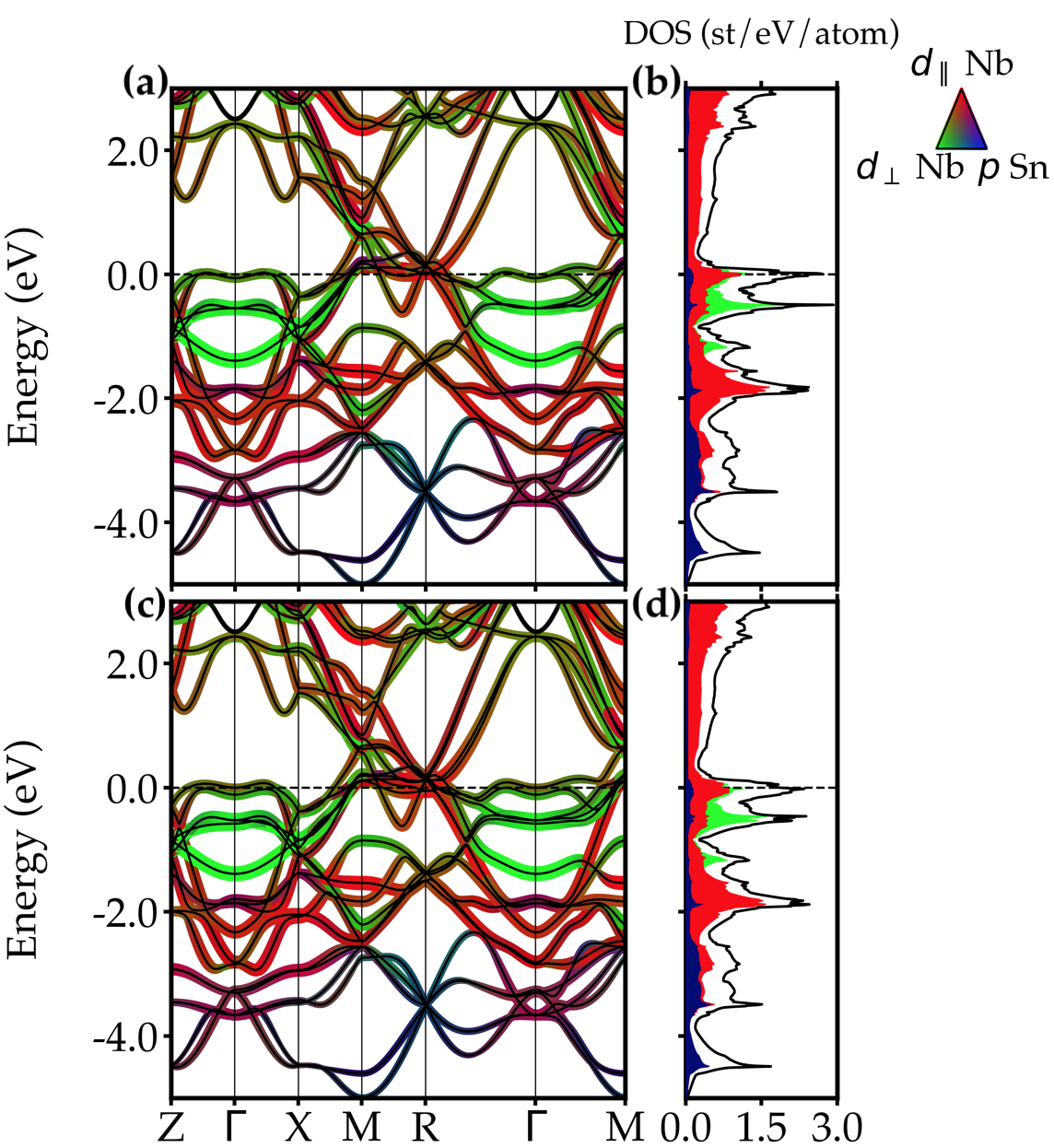}
	\caption{\footnotesize{\textbf{(a-c)} Electronic band structure of cubic and tetragonal Nb$_3$Sn. The thickness of each band reflects the weight of the projected orbitals: Nb-$d_{\parallel}$ (red), Nb-$d_{\perp}$ (green), and Sn-$p$ (blue). Here, $d_{\parallel}$ denotes Nb $d$ orbitals oriented along the direction of the corresponding Nb--Nb chains, while $d_{\perp}$ indicates $d$ orbitals transverse to the chains. \textbf{(b-d)} Orbital-projected density of states (DOS) in units of states/eV/atom. The total DOS is shown in black. The Fermi level is set to zero energy (dashed line).
    }}
    \label{fig:ebands}
\end{figure}

We now discuss in detail the electronic structure of the two phases.
The band structures of cubic and tetragonal Nb$_3$Sn are illustrated in Fig. \ref{fig:ebands}. In panel (a), we report the band structure of the cubic phase, where the thickness of each band indicates the orbital character of the electronic states. We project onto three main components: Nb-$d_{\parallel}$ (red), Nb-$d_{\perp}$ (green), and Sn-$p$ (blue) orbitals. Here, $d_{\parallel}$ is the subset of Nb-$d$ orbitals oriented longitudinally along the corresponding Nb-Nb chain of the A15 structure. $d_{\perp}$ represents transverse $d$ orbitals, i.e. those orthogonal to the chain direction.
\\
\indent
Valence bands extend about 5 eV below the Fermi level. In the energy range from $-5.0$ to $-3.0$ eV, the dominant character is Sn-$p$, with a noticeable contribution from Nb-$s$ states. 
Nb-$d$ orbitals contribute significantly across the entire valence region. Close to the Fermi level, the spectral weight is almost entirely due to Nb-$d$ orbitals, as one can observe from the orbital-projected density of states (DOS) reported in Fig. \ref{fig:ebands} (b). The total DOS exhibits several sharp features, including a prominent peak at the Fermi level, where $N(\varepsilon_F) = 2.4$ st/eV/atom. This peak is caused by flat, nearly dispersionless bands crossing $\varepsilon_F$ in all directions around the $\Gamma$ point and in the M-R directions. These flat bands display a mixture of longitudinal ($d_{\parallel}$) and transverse ($d_{\perp}$) Nb orbitals: around the R point, $d_{\parallel}$ states dominate, while near $\Gamma$, the $d_{\perp}$ component becomes dominant, accounting for up to 75\% of the total orbital weight. Additionally, a flat band along the M-R path also exhibits dominant Sn-$p$ character. A fourfold-degenerate manifold at the R point lies only a few meV above the Fermi level and gives rise to several highly dispersive conduction bands. These bands, which cross $\varepsilon_F$ with large Fermi velocity, are predominantly of $d_{\parallel}$ character. The resulting Fermi surface comprises six sheets (see Fig. S3–S5 in the SM): three tubular sheets along the Brillouin zone edges, reflecting the chain-like $d_{\parallel}$ states from the M–R region, and nearly spherical electron pockets around $\Gamma$, dominated by $d_{\perp}$ orbitals. This orbital decomposition of the Fermi surface will be further discussed in Sec. V.
\\
\indent
In panels (c) and (d) of the same figure, we report for comparison band structure and DOS of the tetragonal phase of Nb$_3$Sn. As reported in earlier works \cite{Sadigh_nb3sn_PRB_1998, DeMarzi_nb3sn_strain_IOP_2013}, the tetragonal strain $\varepsilon = c/a - 1$ does not cause significant changes in the electronic structure. However, the dimerization $\delta$ associated with the $\Gamma_{12}^+$ phonon mode has a much stronger effect. The most noticeable change is the splitting of several flat bands near the Fermi level, particularly those centered around $\Gamma$ and R points. These bands, nearly dispersionless in the cubic phase, lose their degeneracy and shift upwards and downwards in energy in the low-symmetry phase, reducing the DOS at the Fermi level by 16\%. As a consequence, one of the bands becomes fully occupied. This makes the $\Gamma_{12}^+$ phonon potential anharmonic, as the energy gain from band splitting depends nonlinearly on $\delta$ ~\cite{Boeri_PRB_2002_MgB2_anharm}. The magnitude of the anharmonic correction depends on $N(\varepsilon_F)$. This explains why increasing the electronic smearing stabilizes the phase: smearing smooths out sharp features in the DOS, artificially reducing $N(\varepsilon_F)$ and suppressing the instability without addressing its physical origin.

\section{Vibrational properties}

\begin{figure}[t]
  \centering
  \includegraphics[width=0.9\columnwidth]{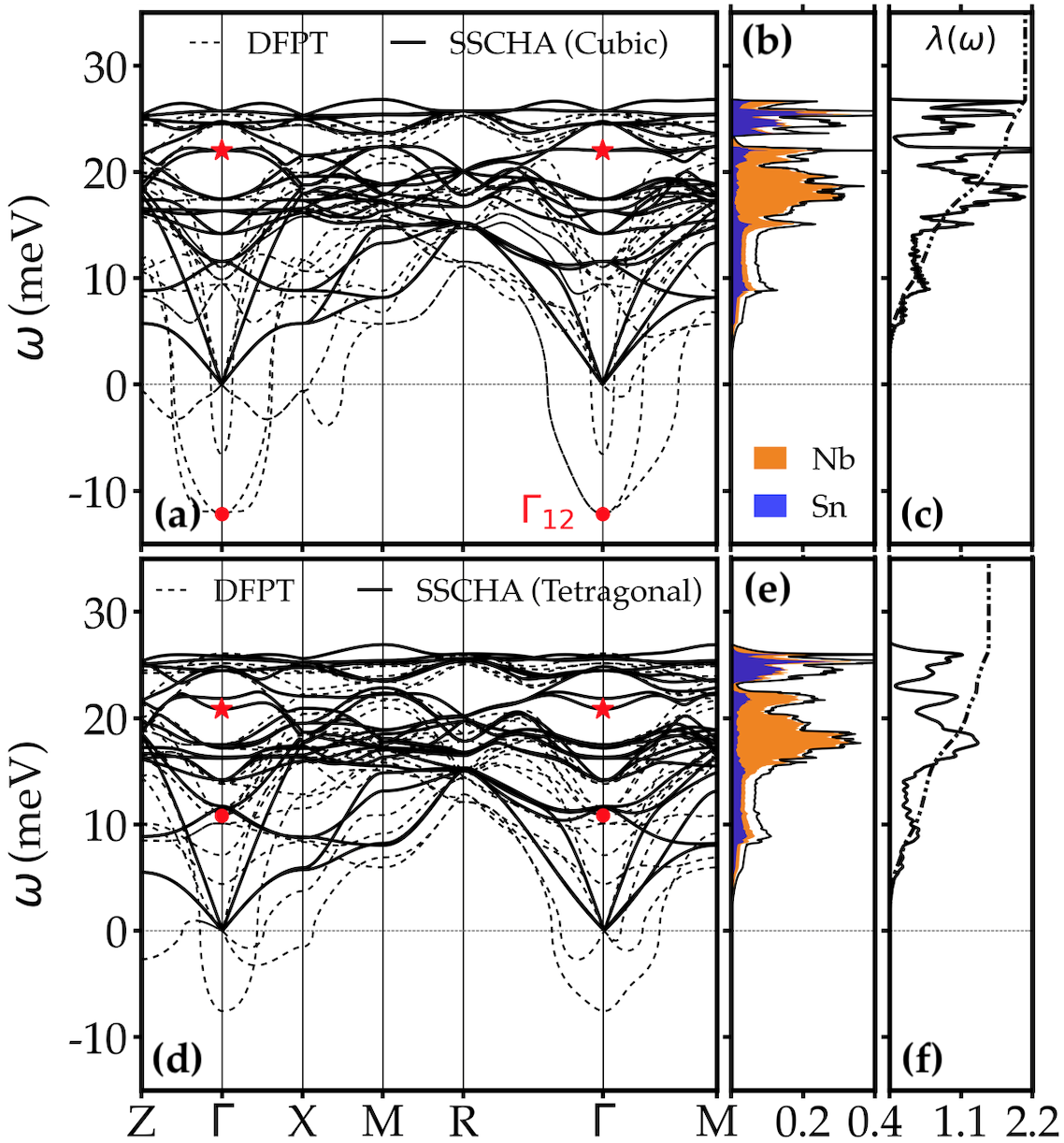}
  \caption{\footnotesize{\textbf{(a-d)} Phonon dispersions computed within harmonic DFPT (black dashed lines) compared to fully anharmonic SSCHA calculations at $T = 0$ K (black solid lines) for the cubic and tetragonal phase of Nb$_3$Sn, respectively. Imaginary phonon modes are represented by negative frequencies. Red circles denote the calculated DFPT frequency of the $\Gamma_{12}^+$ mode, which drives the transition. Red stars denote its renormalization due to anharmonic effects.
  \textbf{(b-e)} Atom-projected and total phonon density of states in units of states/meV. Nb and Sn projections are represented in orange and blue, respectively.
  \textbf{(c-f)} Isotropic Eliashberg spectral functions, along with the electron-phonon coupling $\lambda(\omega)$ (dotted-dashed black line).
  }}
  \label{fig:ph_0}
\end{figure}

\begin{figure*}[t]
	\includegraphics[width=0.8\textwidth]{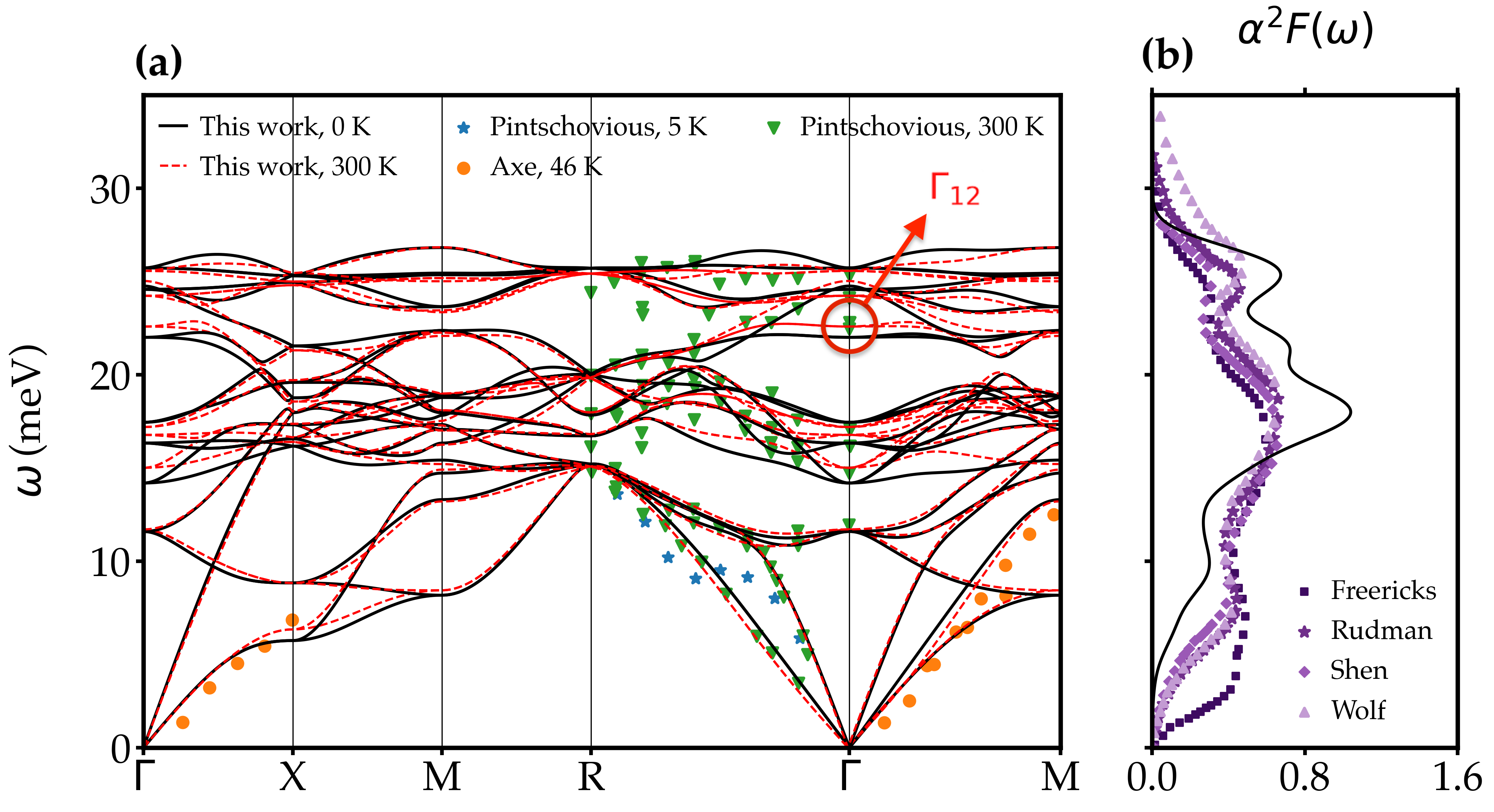}
	\caption{\footnotesize{\textbf{(a)} SSCHA phonon dispersions of cubic Nb$_3$Sn at $T = 0$~K and $T = 300$~K. The results show excellent agreement with inelastic neutron scattering data (colored markers) by Pintschovious \textit{et al.}~\cite{Pintschovius_phonons_nb3sn_PRB_1983, Pintschovius_phonons_nb3sn_PRL_1985} and Axe \textit{et al.}~\cite{Axe_phonons_PRB_1983}. The temperature-dependent renormalization of the $\Gamma_{12}^+$ mode is accurately reproduced, providing a strong validation of our results.
    \textbf{(b)} Calculated Eliashberg spectral function $\alpha^2F(\omega)$ compared with experimental data (purple markers) extracted from tunneling spectroscopy in Refs.~\cite{Shen_nb3sn_a2f_PRL_1972, Rudman_nb3sn_a2f_1984, Freericks_nb3sn_tunnelling_PRB_2002, Kieselmann_nb3sn_tunnelling_1982, Wolf_tunnelling_a2F_2011, Geerk_tunnelling_a15_1985}. The theoretical curve has been broadened for a best comparison to experimental results, which show a significant spread of shapes.}}
	\label{fig:ph_300}
\end{figure*}

We now discuss the vibrational properties of the two phases of Nb$_3$Sn. In panels (a) and (d) of Fig. \ref{fig:ph_0}, we compare the phonon dispersions obtained from harmonic DFPT (dashed lines) and SSCHA (solid lines) at $T = 0$~K. At the harmonic level, both cubic and tetragonal Nb$_3$Sn are dynamically unstable: as expected from the $\epsilon\;\mathrm{vs.}\;\delta$ plots shown in Fig. \ref{fig:BO} (a), the zone-center $\Gamma_{12}^+$ phonon mode (highlighted with red circles) has imaginary frequencies in the cubic phase but real in the tetragonal one, where other instabilities emerge at the X and Z points. Once anharmonic effects are included with SSCHA, all imaginary frequencies disappear and both phases become dynamically stable. The $\Gamma_{12}^+$ mode, in particular, undergoes a strong renormalization (red stars). Taken together, these results support the scenario proposed in Fig.~\ref{fig:BO} (b): the martensitic transition in Nb$_3$Sn is not a Peierls-like second-order transition, driven by a soft mode, but rather a weakly first-order transition between two nearly-degenerate minima, both stabilized by anharmonic effects. Which phase prevails depends on internal residual stress or on deviations from stoichiometry that can lower the cubic minimum compared to the tetragonal one. The atom-projected phonon density of states (DOS) computed with SSCHA are shown in panels (b) and (e) of Fig.~\ref{fig:ph_0}. The vibrational spectrum extends up to $\sim$ 27 meV and can be divided into three regions. The low-frequency part (0–15 meV), associated with acoustic modes, is nearly identical in both phases and shows an equal contribution from Nb and Sn atoms. In the mid-frequency range (15–21 meV), phonon modes related to Nb-chains vibrations, such as the $\Gamma_{12}^+$ phonon, dominate the spectrum. At higher frequencies, both Nb and Sn atoms contribute. 
\\
\indent
In order to validate our SSCHA calculations, we compare the anharmonic phonon dispersions computed with SSCHA at $T = 0$~K and $T = 300$~K for the cubic phase to experimental inelastic neutron scattering data from Axe~\cite{Axe_phonons_PRB_1983} and Pintschovious \cite{Pintschovius_phonons_nb3sn_PRB_1983, Pintschovius_phonons_nb3sn_PRL_1985} -- See Fig.~\ref{fig:ph_300} (a). Our calculations show excellent agreement with both data sets, reproducing not only the energy scale but also the curvature of all major phonon branches. In particular, our calculations accurately reproduce the temperature renormalization of the $\Gamma_{12}^+$ mode, reinforcing the reliability of the SSCHA approach. Minor deviations from the 46~K data by Axe (orange circles) are visible along the $\Gamma$-M path, where an anomalous softening of the longitudinal acoustic branch is experimentally observed \cite{Testardi_review_RMP_1975}. This softening, related to the elastic constant $C_{11}-C_{12}$, is a well known precursor of the martensitic transition, which occurs just above the transition temperature $T_m$ and is restored below, whether or not the transition occurs. Since our calculations are performed at $T=0$~K, they are not expected to reproduce this anomaly, and indeed no such softening is observed, consistent with the rehardening of the elastic constant at low temperature \cite{Testardi_review_RMP_1975}. We also note that our findings do not reproduce the softening of the longitudinal acoustic branch along the $\Gamma$–R direction reported by Pintschovius (blue stars) at low temperature \cite{Pintschovius_phonons_nb3sn_PRL_1985}. This discrepancy is the only notable deviation from experimental data. For reference, we report in Fig.~S2 of the SM the DFPT phonon dispersions of the cubic phase computed using large electronic smearing. While smearing artificially stabilizes the phonons, the resulting spectra deviate significantly from experimental data, confirming that only a proper anharmonic treatment can match experimental data. 

\section{Superconducting properties}

\subsection{Critical temperature}

The isotropic Eliashberg spectral functions of cubic and tetragonal Nb$_3$Sn are shown in panels (c) and (f) of Fig. \ref{fig:ph_0} and compared to experimental data from tunneling spectroscopy in Fig. \ref{fig:ph_300} (b) ~\cite{Shen_nb3sn_a2f_PRL_1972, Rudman_nb3sn_a2f_1984, Freericks_nb3sn_tunnelling_PRB_2002, Kieselmann_nb3sn_tunnelling_1982, Wolf_tunnelling_a2F_2011, Geerk_tunnelling_a15_1985}. The positions of the dominant peaks are in excellent agreement with most references, indicating that the key phonon modes are correctly captured by our calculations. In both cubic and tetragonal Nb$_3$Sn, $\lambda$ increases linearly with phonon frequency, suggesting an even contribution from all phonon branches. The calculated electron-phonon coupling constants and logarithmic average phonon frequencies are reported in Table \ref{tab:tc_table}. 
Our calculated $\lambda = 2.08$ for the cubic phase is slightly larger than the experimental average $\lambda_{\mathrm{exp}} = 1.83$ \cite{Mentink_nb3sn_IOP_2017}, but determining
$\lambda$ in Nb$_3$Sn is notoriously challenging, both experimentally and computationally \cite{Kieselmann_nb3sn_tunnelling_1982, Freericks_nb3sn_tunnelling_PRB_2002, Wolf_tunnelling_a2F_2011, Lucrezi_FBW_2024}, due to the strong energy dependence of the electronic DOS near a van Hove singularity (VHS). Indeed, the VHS makes all quantities extremely noisy without a full-bandwidth (FBW) approach \cite{Lucrezi_FBW_2024}. 
\\
\indent
In order to reduce the effect of noise on the superconducting properties, we solved the anisotropic Migdal–Eliashberg equations within the FBW framework, as recently implemented in the EPW code \cite{Lee_NPJ_epw_2023}. Results are summarized in Table \ref{tab:tc_table}, where they are compared to experimental data, if available. The Coulomb interaction is treated within the Morel–Anderson pseudopotential framework~\cite{Morel_PhysRev_1962_mustar}, using $\mu^*$ values for the cubic phase derived from first-principles Random Phase Approximation (RPA) and Kukkonen–Overhauser (KO) calculations by Pellegrini \textit{et al.}~\cite{Pellegrini_Nb_PRB_2023}. Values for the tetragonal phase were obtained rescaling the cubic $\mu^*$ by the DOS value -- See Table S2 in SM.
\\
\indent
The superconducting gap $\Delta_{\mathbf{k}}$ of the two phases, shown in Fig.~\ref{fig:gap} (a), displays a broad but continuous anisotropic distribution over the Fermi surface. In the cubic phase, the largest gap values (6.3 meV) are found near the R points. Here the Fermi velocity is low and electronic states are dominated by longitudinal $d_{\parallel}$ orbitals of the Nb chains -- See Fig.~S3-4 of the SM. In contrast, the smallest gap values (3.1 meV) appear on the electron pockets near $\Gamma$, where the Fermi velocity is low but electron-phonon coupling is weaker. As discussed in Sec. III, these pockets are mainly composed of transverse $d_{\perp}$ states from the Nb chains. The orbital-distribution of the gap challenges the traditional view that superconductivity in Nb$_3$Sn is driven by one-dimensional Nb chains. Instead, it supports a three-dimensional pairing mechanism that involves both longitudinal and transverse Nb orbitals. If pairing had involved only longitudinal states, the gap would be highly directional, with nodes or minima corresponding to off-chain states. Instead, the gap is broadly anisotropic but remains fully-open, in agreement with most recent experiments. These findings also help explain why $T_c$ is suppressed in off-stoichiometric or disordered samples, even when Nb chains remain structurally intact~\cite{Charlesworth_nb-sn_1970, Godeke_review_IOP_2006, Mentink_nb3sn_IOP_2017}: transverse $d_\perp$ states are particularly sensitive to local changes in the Sn environment, making Sn stoichiometry a key tuning parameter. 
\\
\indent
After the martensitic transition takes place, the Fermi surface is slightly reshaped with noticeable changes in both Fermi velocity and gap distributions. In particular, the high-gap sheets near the R points lose spectral weight, while the two portions of FS coming from the electron pockets around $\Gamma$ show an increased Fermi velocity and reduced gap in the tetragonal direction around the Z points. 
\\
\indent
The temperature dependence of the superconducting gaps is shown in Fig. \ref{fig:gap} (b). Both isotropic and anisotropic solutions yield nearly identical critical temperatures for the two phases. 
Using a value of $\mu^*$ extracted from simple RPA yields a too high $T_c$ for the cubic phase ($T_c=25$ K) compared to the experimental value ($T^{\,\mathrm{exp}}_c = 18.3$ K). From KO, we obtain a theoretical $T_c =22$ K for the cubic phase and 16 K for the tetragonal phase, in much closer agreement with experiments. While the cubic $T_c$ is still slightly overestimated, the error is within the typical 10–20\% uncertainty of \textit{ab-initio} methods \cite{Sanna_PRL_2020_SCDFT_bench} and the calculated BCS ratios $2\Delta/k_B T_c$, reported in Table \ref{tab:tc_table}, fall nicely within the experimental range 4.2-4.9. 

\subsection{Critical fields}

\begin{table*}[t] 
\centering 
\renewcommand{\arraystretch}{1.5} 
\setlength{\tabcolsep}{3.2pt} 
\begin{tabular}{lcccccccccccccc} 
\hline 
\hline 
Phase & $N(\varepsilon_F)$ & $\lambda$ & $\omega_{\log}$ & $\langle v_F^* \rangle$ & $\langle \Delta \rangle$ & $T_c$ & $T_c^{\,\exp}$ & $\frac{2\Delta}{k_B T_c}$ & $\langle \xi_0 \rangle$ & $\xi_{\mathrm{GL}}^{\mathrm{\,c}}(0)$ & $\ell^{\,\mathrm{exp}}$ & $\xi^{\exp}_{\mathrm{GL}}(0)$ & $H_{c2}^{\,\mathrm{c}}(0)$ & $H_{c2}^{\exp}(0)$ \\ 
 & (st/eV/at) & & (meV) & ($10^7$cm/s) & (meV) & (K) & (K) & & (nm) & (nm) & (nm) & (nm) & (T) & (T) \\ 
\hline 
Cubic   & 2.4 & 2.08 & 15.2 & 0.71 & 4.4 & 22 & 18 & 4.7 & 3.9 & 3.1 & 2.6-3.3 & 3.3 & 32 & 29 \\ 
Tetrag. & 2.0 & 1.52 & 13.1 & 0.87 & 3.1 & 16 & 17 & 4.4 & 6.6 & 4.9 & 10.0 & 4.0 & 14 & 21 \\
Cubic*   & -- & -- & -- & -- & 3.7 & 18 & 18 & 4.7 & 4.8 & 3.6 & -- & 3.3 & 26 & 29 \\
\hline 
\hline 
\end{tabular} 
\caption{Calculated superconducting properties of cubic and tetragonal Nb$_3$Sn. 
$N(\varepsilon_F)$ is the electronic DOS at the Fermi level; 
$\lambda$ is the electron-phonon coupling constant; 
$\omega_{\log}$ is the logarithmically averaged phonon frequency; 
$\langle v_F \rangle$ is the average Fermi velocity over the FS; 
$\langle \Delta \rangle$ is the average superconducting gap; 
$T_c$ and $T_c^{\,\exp}$ are the theoretical and experimental critical temperatures, respectively; 
$2\Delta/k_B T_c$ is the BCS ratio; 
$\langle \xi_0 \rangle$ is the average Pippard coherence length; 
$\xi_{\mathrm{GL}}^{\,\mathrm{c}}(0)$ is the GL coherence length calculated at $T=0$~K in the clean limit; 
$\ell^{\,\mathrm{exp}}$ is the experimental mean free path; 
$\xi^{\,\exp}_{\mathrm{GL}}(0)$ is the experimental GL coherence length; 
$H_{c2}^{\mathrm{c}}(0)$ is the upper critical field at $T=0$~K, calculated in the clean limit and neglecting Pauli limiting; 
$H_{c2}^{\exp}(0)$ is the experimental upper critical field measured in nearly stoichiometric samples. Cubic* reports the same calculations as for the cubic phase, but after rescaling the superconducting gap to match experimental $T_c$. Experimental data are taken from Refs. \cite{Orlando_nb3sn_PRB_1979, Mentink_nb3sn_IOP_2017}.}
\label{tab:tc_table} 
\end{table*}

\begin{figure}[b]
\includegraphics[width=0.90\columnwidth]{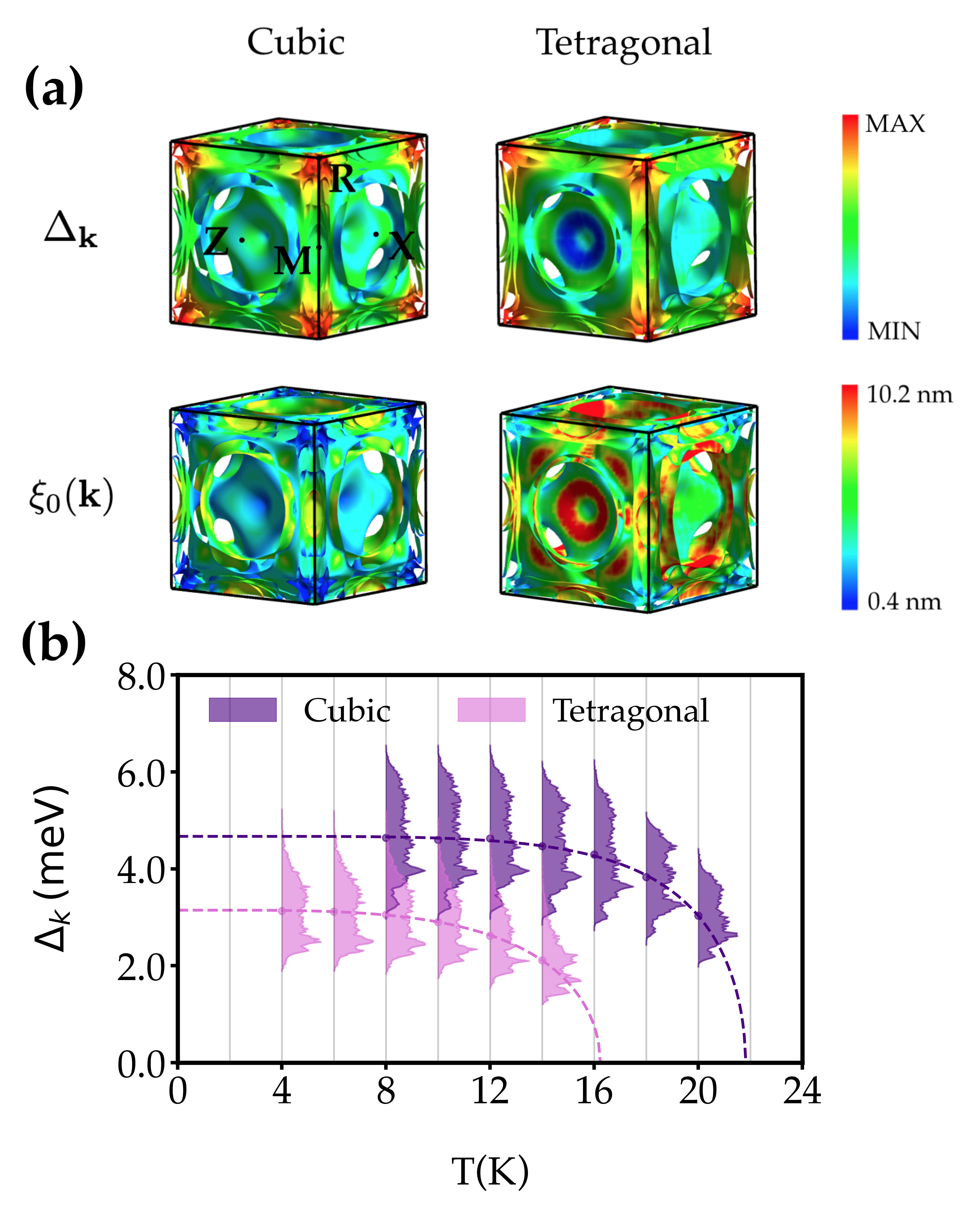}
    \caption{\footnotesize{\textbf{(a)} Distribution of the superconducting gap (top) and Pippard coherence length (bottom) over the Fermi surface at 8 K for the cubic (left) and tetragonal (right) phases of Nb$_3$Sn. The color scale indicates the magnitude of the superconducting gap $\Delta_{\mathbf{k}}$, ranging from its minimum (blue) to maximum (red) value across the Fermi surface, and the magnitude of the coherence length $\xi_0({\mathbf{k}})$ ranging from the minimum (0.4 nm) to the maximum (10.2 nm) value of the cubic phase.
    \textbf{(b)} Energy distribution of the zero-frequency superconducting gap of Nb$_3$Sn as a function of temperature, obtained solving the anisotropic Migdal–Eliashberg equations using the FBW approach \cite{Lucrezi_FBW_2024}. Dashed lines indicate a fit of the weighted averages of the distribution for each T.}}
    \label{fig:gap}
\end{figure}

Having access to the momentum-dependent solution of the ME equations, we can compute the momentum-resolved Pippard coherence length:

\begin{equation}\label{pippard}
    \xi_0({\mathbf{k}}) = \frac{\hbar v_F(\mathbf{k})}{\pi Z_\mathbf{k} \Delta_\mathbf{k}}, \qquad Z_\mathbf{k}=1+\lambda_\mathbf{k}
\end{equation}
\\
\noindent
In this expression, the Fermi velocities are renormalized by the normal state self-energy $Z_\mathbf{k}$ to account for strong-coupling corrections \cite{Orlando_nb3sn_PRB_1979}.
The distribution of $\xi_0(\mathbf{k})$ over the Fermi surface is shown in Fig.~\ref{fig:gap} (a), together with that of the superconducting gap. We find that the anisotropy of $\xi_0(\mathbf{k})$ is mainly controlled by the renormalized Fermi velocity $v_F^*=v_F/Z$, rather than directly by the gap: both low and high-gap regions give rise to short $\xi_0$ if $v_F^*(\mathbf{k})$ is small. In the cubic phase, the electron pockets near $\Gamma$ are short-$\xi_0$ regions due to their low $v_F(\mathbf{k})$ -- see also Fig.~S5 of the SM. After the martensitic transition, these pockets become large-$\xi_0$ regions: along the tetragonal Z direction, $v_F(\mathbf{k})$ increases while the $\lambda_{\mathbf{k}}$ and $\Delta_{\mathbf{k}}$ decrease. 
\\
\indent
Using the calculated $\xi_0$, we can obtain a fully \textit{ab initio} estimate of the zero-temperature upper critical field $H_{c2}(T=0)$. This is done by averaging $\xi_0(\mathbf{k})$ over the Fermi surface to extract an effective Ginzburg–Landau coherence length $\xi_{\mathrm{GL}}(0)$. The upper critical field then follows as:

\begin{equation}\label{hc2_GL}
   H_{c2}(0)
   =
   \frac{\phi_{0}}{2\pi\,\xi_{\mathrm{GL}}^{2}(0)}\,, \:\quad \phi_0=2.068\times10^{-15}\;\mathrm{T\cdot m}^2
\end{equation}
\\
\noindent
where $\phi_0$ is the magnetic flux quantum. In the clean limit, $\xi_{\mathrm{GL}}^\mathrm{\,c}(0)\approx 0.74\,\xi_0$ \cite{Helfand_hc2_1966}, which implies $H_{c2}^\mathrm{\,c}(0)\propto 1/\xi_0^2$. Hence, an increase in the average coherence length reduces the upper critical field. We find that $\langle \xi_0 \rangle$ increases from 3.9 nm in the cubic phase to 6.6 nm in the tetragonal phase, consistent with the suppression of $H_{c2}$ observed in experiments with the martensitic transition. Our clean limit estimates are $H_{c2}^{\mathrm{c}}(0)=32$~T and 14 T for the cubic and tetragonal phase, respectively. However, the perfectly-clean limit corresponds to $\xi_0 \ll \ell$, while typical Nb$_3$Sn samples lie in the intermediate-to-dirty domain ($\xi_0 \sim \ell$), with mean free paths $\ell^{\,\mathrm{exp}}$ ranging from 2.6 nm in cubic samples to 10 nm in tetragonal ones \cite{Orlando_nb3sn_PRB_1979, Mentink_nb3sn_IOP_2017}. Thus, our clean limit estimates should be regarded as lower bounds for $H_{c2}(0)$. The calculated upper critical field for the cubic phase exceeds the experimental value $H^{\mathrm{exp}}_{c2}(0)=29$~T by $\sim10\%$. However, we observe that also $T_c$ -- and hence $\Delta$ -- are larger than in experiments. If we rescale the gap and the $T_c$ value, keeping $2\Delta/k_B T_c$ fixed (Cubic* in Table \ref{tab:tc_table}), the estimate improves to 26 T, which now lies below experiment and represents a consistent lower bound. For the tetragonal phase, on the other hand, the clean limit prediction of 14 T underestimates the experimental 21 T by $\sim50\%$ \cite{xi_footnote}.
Thus, we find that clean limit formulas work reasonably well for samples in the clean-to-intermediate limit, where $H_{c2}$ is still dominated by Fermi surface effects. This allows us to also speculate about possible mechanism to tune the superconducting properties of A15 Nb$_3$Sn samples.
\\
\indent
In particular, based on our findings, we can provide a clear picture of how different doping strategies affect superconducting performance. Substitutional dopants introduce additional scattering, which drives the system toward the dirty limit and thereby increases $H_{c2}$ while enhancing flux pinning and $J_c$. Beyond this common effect, the impact depends on which electronic states are modified. Doping on Sn-sites selectively enhances the electron–phonon coupling of transverse $d_{\perp}$ states, which correspond to low-gap, large-$\xi_0$ regions. Strengthening these states reduces anisotropy and simultaneously increases both $T_c$ and $H_{c2}$. This is consistent with reports of Al doping, which above 6\% suppresses the structural transition and increases $T_c$ up to 18.5 K, the highest value observed in Nb$_3$Sn \cite{Vieland1970TcReport}. In contrast, doping on the Nb-sites (e.g., with Ti, Ta, or Hf) primarily affects the longitudinal Nb-$d$ states. This impacts the same bands that form the high-gap pockets at R, but also Fermi surface sheets where coherence lengths are largest. Because these extended regions limit $H_{c2}$, Nb-sites doping can help reduce coherence lengths and make $H_{c2}$ more robust. However, since it also weakens the strongly coupled chain states, it tends to suppress $T_c$ \cite{Heald_nb3sn_tita_2018}.

\section{Conclusions}

In this work, we presented the first fully \textit{ab initio} microscopic description of the superconducting state of Nb$_3$Sn, one of the workhorse superconductors for high-field applications. Our goal was to provide a reliable theoretical foundation for understanding and optimizing its performance, something that had remained out of reach for many years.
\\
\indent
Our results address three key open questions:
i) the martensitic transition in Nb$_3$Sn is not a second-order Peierls instability, but a weakly first-order transition between two anharmonically stabilized structures. The tetragonal phase is never the global minimum of the potential energy surface in our calculations, but may be stabilized in real samples by internal stress;
ii) the superconducting gap is single-valued but strongly anisotropic, in line with recent specific heat measurements \cite{Escudero_nb3sn_firstorder_2009,Jo_nb3sn_gap_2014};
iii) the martensitic transition does suppress the upper critical field $H_{c2}$, not simply because of the reduced coupling, but due to a redistribution of coherence lengths across the Fermi surface.
\\
\indent
Beyond answering these questions, we uncovered several new insights. We showed that accurately describing Nb$_3$Sn requires going beyond standard DFT approximations. Harmonic phonon calculations -- even with numerical artifacts like increasing electronic smearing -- fail to reproduce experimental phonon spectra. Anharmonicity, as observed for NbTi and NbN \cite{Cucciari_nbti_PRB_2024, Kogler_nbn_arxiv}, appears to be a general feature of real-world materials which is essential to stabilize both the cubic and tetragonal structures. In addition, reproducing the experimental $T_c$'s requires an advanced treatment of the Coulomb interaction beyond the empirical $\mu^*$ approximation and a full-bandwidth solution of the Migdal–Eliashberg equations. 
This work would not have been possible without recent community efforts to extend \textit{ab initio} methods to real-world superconductors \cite{Tresca_PRB_2022, Sitaraman_nbzr_2023, DiCataldo_nbc_arxiv_2025, Kogler_nbn_arxiv, Jones_nbti_fs_prb_2025}.
\\
\indent
Finally, we showed that the anisotropy of the superconducting gap arises from an unexpected three-dimensional pairing mechanism. Contrary to the traditional view \cite{Weber_nb3sn_bands_PRB_1982}, pairing is not limited to longitudinal $d_{\parallel}$ orbitals along the Nb chains, but also involves transverse $d_{\perp}$ orbitals with nearly equal weight. This leads to a strongly anisotropic single gap, consistent with experiment. The resulting coherence length $\xi_0({\mathbf{k}})$ is primarily controlled by the renormalized Fermi velocity rather than by the gap amplitude. The martensitic transition reshapes the Fermi surface such that the $\Gamma$-centered electron pockets, which are short-$\xi_0$ in the cubic phase, become long-$\xi_0$ in the tetragonal one. This redistribution of coherence lengths increases the average $\xi_0$ and provides a microscopic explanation for the experimentally observed drop of $H_{c2}$ across the martensitic transition.
\\
\indent
These findings suggest practical strategies for optimizing Nb$_3$Sn performance. Doping on Sn-sites may enhance the coupling of transverse orbitals and raise both $T_c$ and $H_{c2}$, while also reducing gap anisotropy.
Al doping is a promising candidate, having already shown the highest $T_c$ in Nb$_3$Sn to date. Nb-sites are usually doped with Ti, Ta or Hf to prevent the martensitic transition and improve $H_{c2}$, despite lowering $T_c$ \cite{Heald_nb3sn_tita_2018}. Our results help explain this trade-off: doping directly on Nb-sites affects the bands responsible for the long-$\xi_0$, reinforcing $H_{c2}$, but may also weaken the large-gap pockets at R, leading to a drop in $T_c$. Still, even without chemical doping, it may be possible to prevent the martensitic transition by minimizing internal stress -- for example, through slow annealing or epitaxial growth --- thus maintaining higher superconducting performance. Although these strategies are rarely pursued, due to the difficulty of controlling Sn diffusion, epitaxial growth and thin-film deposition are actively studied in the context of superconducting radio-frequency (SRF) cavities, where they have already shown excellent results.


\section{Methods}

Electronic and vibrational properties were computed using Density Functional Perturbation Theory (DFPT) within a plane-wave pseudopotential framework, as implemented in the \textsc{Quantum ESPRESSO} suite \cite{Baroni_RevModPhys_2001_DFPT, quantumespresso_1}. The plane-wave basis set employed a kinetic energy cutoff of 80 Ry, ensuring convergence of the total energy to within 1 meV/atom. Scalar-relativistic Optimized Norm-Conserving Vanderbilt (ONCV) pseudopotentials \cite{Hamann_PRB_2017_ONCV} were used in conjunction with the PBEsol exchange-correlation functional \cite{Perdew_pbesol_PRL_2008} for a better agreement with experimental data. The structures were pre-relaxed using and then relaxed within SSCHA. Structural relaxations were carried out until the residual atomic forces were smaller than 2 meV/\r{A}. Brillouin zone integrations were performed on a $12 \times 12 \times 12$ $\Gamma$-centered Monkhorst-Pack $k$-mesh~\cite{Monkhorst_PRB_1976} using Methfessel-Paxton smearing of 0.005 Ry~\cite{Methfessel_PRB_1989}, and a $6 \times 6 \times 6$ mesh for phonons.
\\
\indent
To account for anharmonic effects in lattice dynamics, we employed the Stochastic Self-Consistent Harmonic Approximation (SSCHA), as implemented in the \textsc{SSCHA} Python package \cite{Errea_PRB_sscha_2014, Monacelli_JPCM_2021_SSCHA}. The initial guess for the force-constant matrix $\Phi$ was derived from DFPT dynamical matrices computed on $2^3$, $4^3$, and $6^3$ $q$-grids. The SSCHA minimization starts with a population of 200 individuals. The minimization process is then carried out using larger populations of 400 individuals. A final population of 13.000 individuals is used to extract the hessian,
which is more sensible to the number of configurations. The anharmonic phonon dispersions are obtained from the positional free-energy Hessians without the fourth-order term.
\\
\indent
To access larger supercells and ensure convergence, we used Moment Tensor Potentials (MTPs) \cite{Novoselov_MTP_2019}, trained and evaluated using the \textsc{MLIP} package \cite{Lucrezi_MLIP_2023, Novikov_MLIP_2021, Deringer_MLIP_2019}.  In order to train the potential, we generated a training set of 280 configurations and a validation set of 120 $2\times2\times2$ supercell configurations of the tetragonal phase, using
the SSCHA code. To broaden the configurational sampling, the SSCHA temperature was set to 300 K during training. The MTP used 8 radial basis functions with Chebyshev polynomials defined over the range [2.0 \r{A}, 5.0 \r{A}]. The potential accuracy was assessed via the root mean square error (RMSE) in energy with respect to DFT reference calculations on the validation set. After hyperparameter tuning, a level-24 potential was selected, yielding an RMSE of 0.3 meV/atom.
\\
\indent
To evaluate the superconducting properties of Nb$_3$Sn, we combined the fully anharmonic phonon spectra obtained from SSCHA with electron-phonon matrix elements computed via DFPT. These matrix elements were calculated on coarse $\textbf{q}$- and $\textbf{k}$-point grids and interpolated onto dense meshes using Wannier functions, following the EPW methodology~\cite{Lee_NPJ_epw_2023}. We interpolated to $36^3/72^3$ and $18^3/52^3$ fine meshes for the cubic and tetragonal phase, respectively. 
\\
\indent
\section{Acknowledgments}
The authors thank A. Sanna for discussion about upper critical field calculations.
The authors acknowledge computational resources from the Vienna Scientific Cluster, project 71754 "TEST". A.C. acknowledges funding from Fondazione Sapienza, Progetto Avvio alla Ricerca n. AR124190789AE8E422.
L.B. acknowledges support from
Fondo Ateneo Sapienza 2019-22, and funding from the European Union - NextGenerationEU under the Italian Ministry of University and Research (MUR), “Network 4 Energy Sustainable Transition - NEST” project (MIUR project code PE000021, Concession Degree No. 1561 of October 11, 2022) - CUP C93C22005230007.
\bibliographystyle{apsrev4-1}
\bibliography{library}

\begin{thebibliography}{78}
\bibitem [{Hyb()}]{Hybrid_functional_footnote}%
  \BibitemOpen
  \href@noop {} {}\bibinfo {note} {It is worth noting that the energy
  differences involved are so small that we cannot exclude other effects could
  reshape the BO surface and stabilize a tetragonal phase. Indeed, test
  calculations using the PBE0 hybrid functional [79] predict
  a larger energy gain ($\sim$5 meV/atom) for the experimental tetragonal
  structure over the cubic phase, at least at the harmonic level.}\BibitemShut
  {Stop}%
\bibitem [{xi_()}]{xi_footnote}%
  \BibitemOpen
  \href@noop {} {}\bibinfo {note} {Our calculations refer to ideal
  stoichiometric crystals and do not include extrinsic effects such as
  inhomogeneity or residual cubic domains, which may justify why the
  experimentally observed $T_c$ reduction (typically ~1 K) is much smaller than
  the correspondent $H_{c2}$ reduction.}\BibitemShut {Stop}%
\end{thebibliography}
\end{document}


\author{Alessio Cucciari}\email{alessio.cucciari@uniroma1.it}
\affiliation{Dipartimento di Fisica, Sapienza - Universit\`a di Roma, 00185 Rome, Italy}
\author{Lilia Boeri} \email{lilia.boeri@uniroma1.it}
\affiliation{Dipartimento di Fisica, Sapienza - Universit\`a di Roma, 00185 Rome, Italy}

\title{Supplementary Material for: An \textit{ab initio} answer to long-standing questions about superconducting \ce{Nb3Sn}}
\date{\today}
\maketitle
\noindent
In this Supplementary Material we provide additional details and figures about the methods employed in the main text.
\\
\indent
Section~\ref{sect:Details_abinitio} outlines the computational parameters used in our DFT and DFPT calculations. In Section~\ref{sect:PES_smear}, we show how increasing the electronic smearing in DFT calculations affects the shape of the Born-Oppheneimer potential energy surface. In Section~\ref{sect:Band_structures}, we compare the electronic band structures and density of states of the cubic and tetragonal phases, while Section~\ref{sect:Band_band_fermisurface} presents a band-by-band analysis of the Fermi surface, including orbital character, superconducting gap, Fermi velocities and coherence lengths. Section \ref{sect:Ph_smear} shows how increasing the electronic smearing affects the phonon calculations. Details on the SSCHA optimization and its convergence are discussed in Section~\ref{sect:SSCHA_details}, while Section~\ref{sect:Mlip_details} describes the training and validation of machine-learned interatomic potentials used to accelerate the SSCHA calculations. Section~\ref{sect:Solving_ME} reports details about the solution of the full-bandwidth anisotropic Migdal-Eliashberg equations, for instance how we derived $\mu^*$ from the bare Coulomb interaction computed within RPA and Kukkonen-Overhauser. Finally, convergence tests for DFPT phonons are shown in Section~\ref{sect:conv_test}.

\section{Details of ab-initio calculations}
\label{sect:Details_abinitio}
Electronic and vibrational properties were computed within Density Functional Perturbation Theory (DFPT) in a plane-wave pseudopotential framework, as implemented in the Quantum ESPRESSO (QE) suite \cite{Baroni_RevModPhys_2001_DFPT, quantumespresso_1}. The wave functions expansion was performed with a kinetic energy cutoff of 80 Ry, achieving a convergence within 1 meV/atom on the total energy. We employed the scalar-relativistic version of Optimized Norm-conserving Vanderbilt (ONCV) pseudopotentials \cite{Hamann_PRB_2017_ONCV}, with a Perdew-Burke-Ernzerhof (PBE) exchange-correlation functional \cite{Perdew_PRL_1996_PBE}. Structures were relaxed in QE until each component of the forces acting on single atoms was less than 2.0 meV/Å$^3$. The integration over the Brillouin zone was carried out using a regular $14\times14\times14$ $\Gamma$-centered Monkhorst-Pack \cite{Monkhorst_PRB_1976} grid for electrons, with a Methfessel-Paxton smearing of width 0.005 Ry \cite{Methfessel_PRB_1989}, and a $6\times6\times6$ mesh for phonons. These meshes were chosen in order to achieve a reasonable convergence of the zone-center $\Gamma_{12}^+$  optical mode - see convergence tests plot in Sect. \ref{sect:conv_test}.



\newpage

\section{Effect of smearing on BO potential energy surface}
\label{sect:PES_smear}
\vskip10mm
\begin{figure}[ht]
    \centering
	\includegraphics[width=0.6\columnwidth]{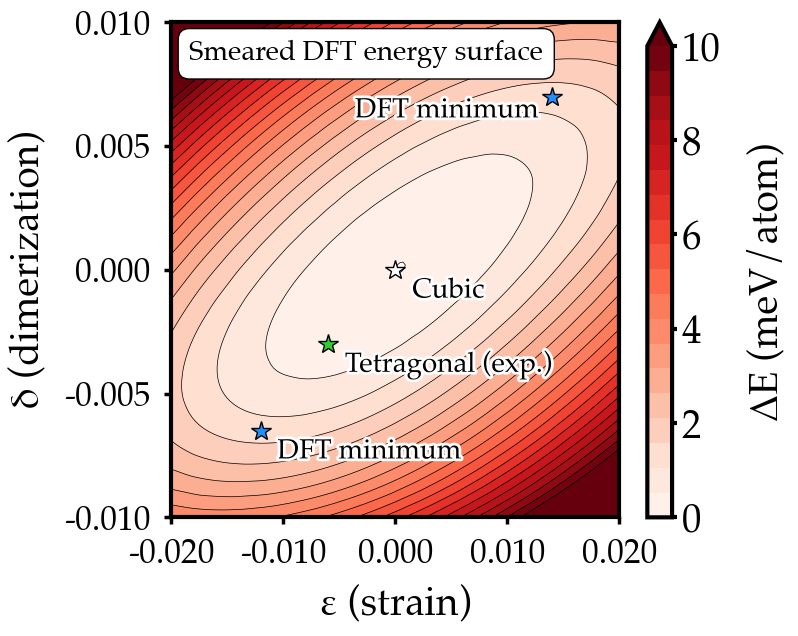}
    \caption{\footnotesize{ Contour plot of the fixed-volume Born–Oppenheimer (BO) Potential Energy Surface $\Delta E(\varepsilon, \delta)$ computed with DFT using an increased Methfessel–Paxton electronic smearing of 0.02 Ry \cite{Methfessel_PRB_1989}.}}
	\label{suppfig:pes_smearing}
\end{figure}

\newpage

\section{Cubic vs Tetragonal band structure}
\label{sect:Band_structures}
\vskip10mm
\begin{figure}[ht]
    \centering
	\includegraphics[width=0.9\columnwidth]{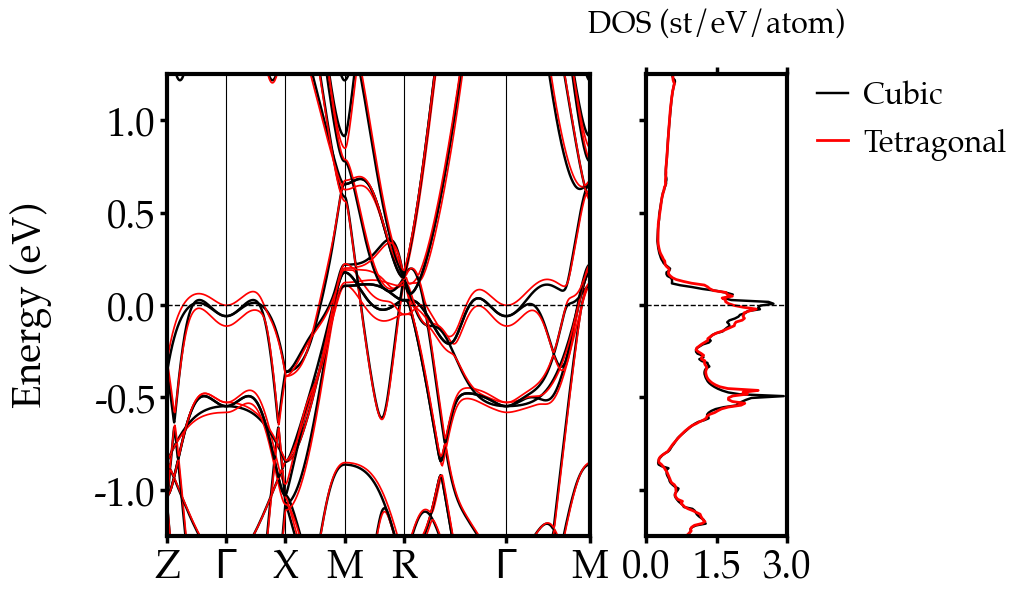}
	\caption{Comparison between the electronic band structures and DOS of the cubic (black solid lines) and tetragonal (red solid lines) phases of \ce{Nb3Sn}.}
	\label{suppfig:band_comp}
\end{figure}

\newpage

\section{Band-by-band decomposition of the Fermi surface}
\label{sect:Band_band_fermisurface}
\vskip10mm
\begin{figure}[ht]
    \centering
	\includegraphics[width=0.75\columnwidth]{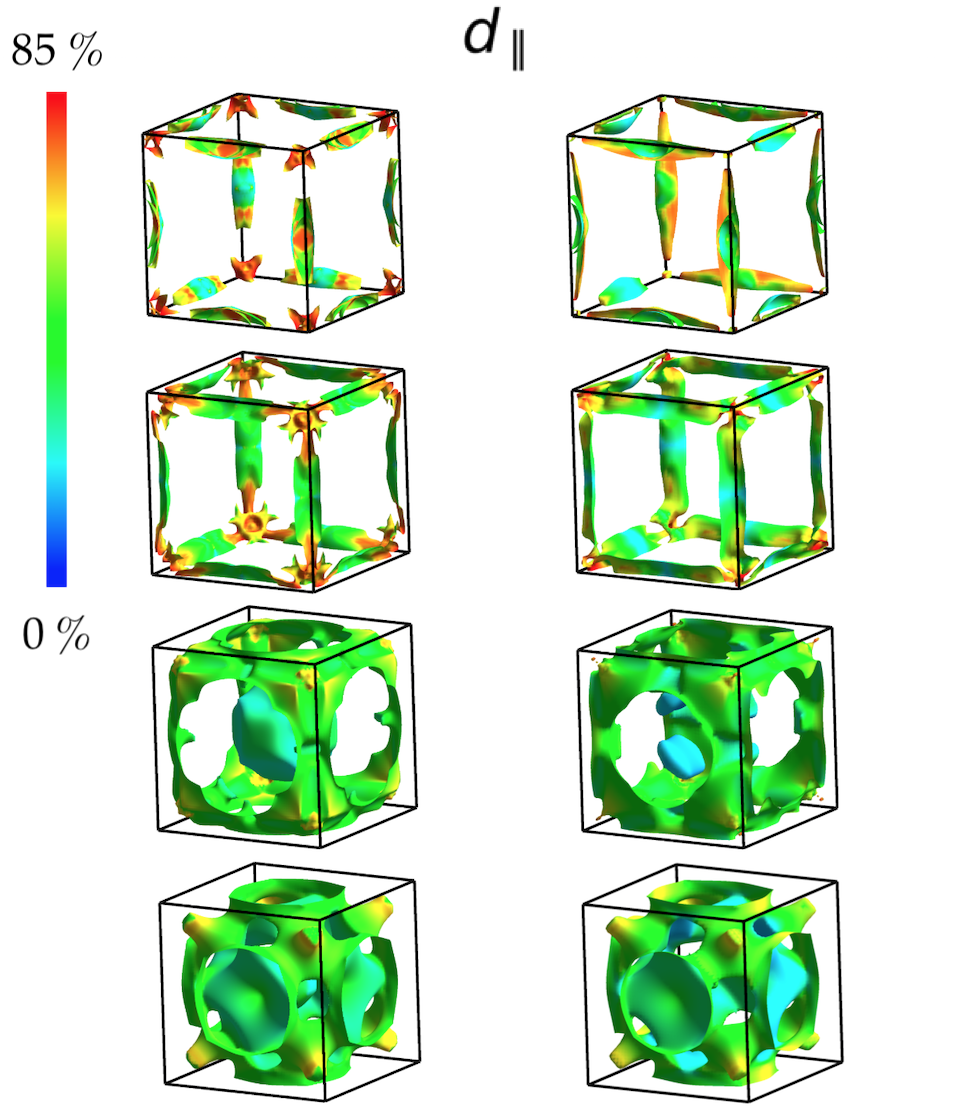}
	\caption{Fermi surface of cubic and tetragonal \ce{Nb3Sn}, broken down into single bands. Each band is decorated with \ce{Nb}-$d_{\parallel}$ orbital character. The color scale goes from zero to the maximum value of the projection in the cubic phase.}
	\label{suppfig:FSlong}
\end{figure}

\newpage

\begin{figure}[ht]
    \centering
	\includegraphics[width=0.75\columnwidth]{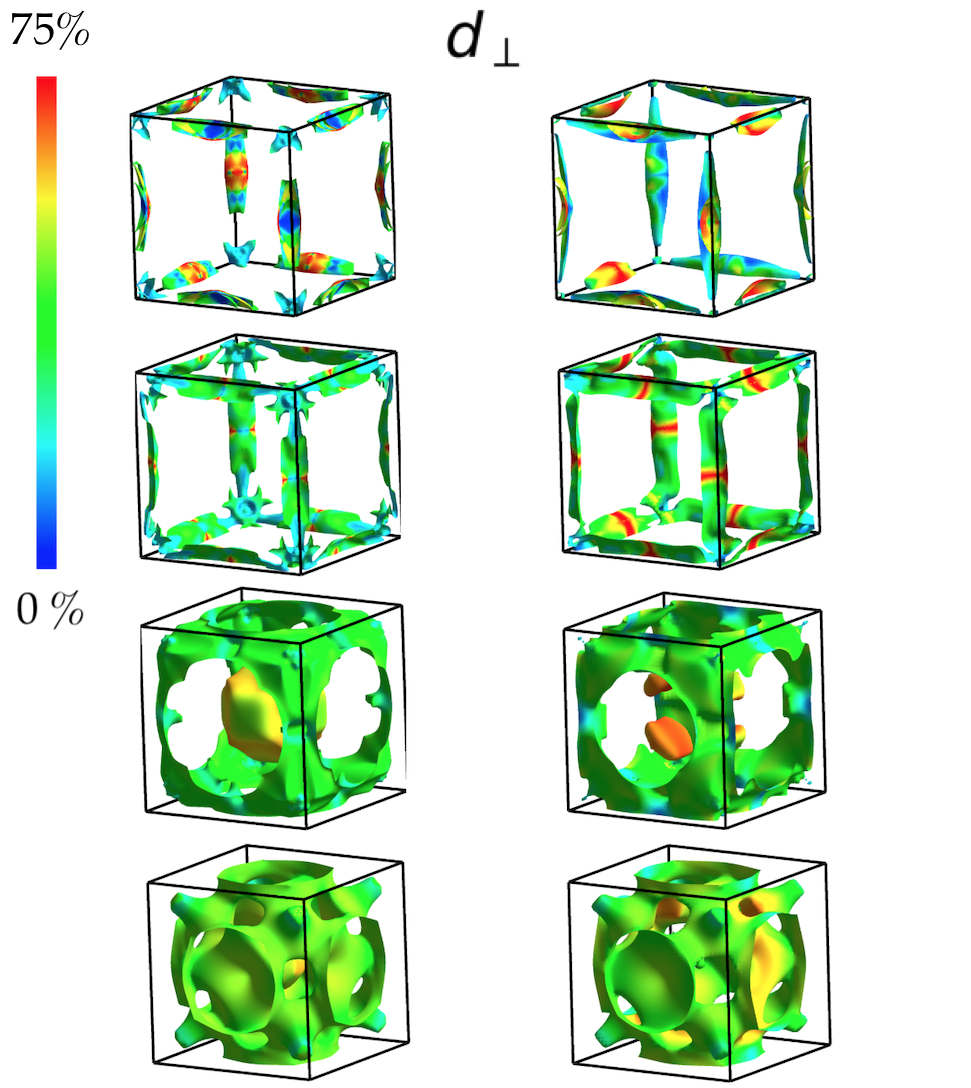}
	\caption{Fermi surface of cubic and tetragonal \ce{Nb3Sn}, broken down into single bands. Each band is decorated with \ce{Nb}-$d_{\perp}$ orbital character. The color scale goes from zero to the maximum value of the projection in the cubic phase.}
	\label{suppfig:FStransv}
\end{figure}

\newpage

\begin{figure}[ht]
    \centering
	\includegraphics[width=0.75\columnwidth]{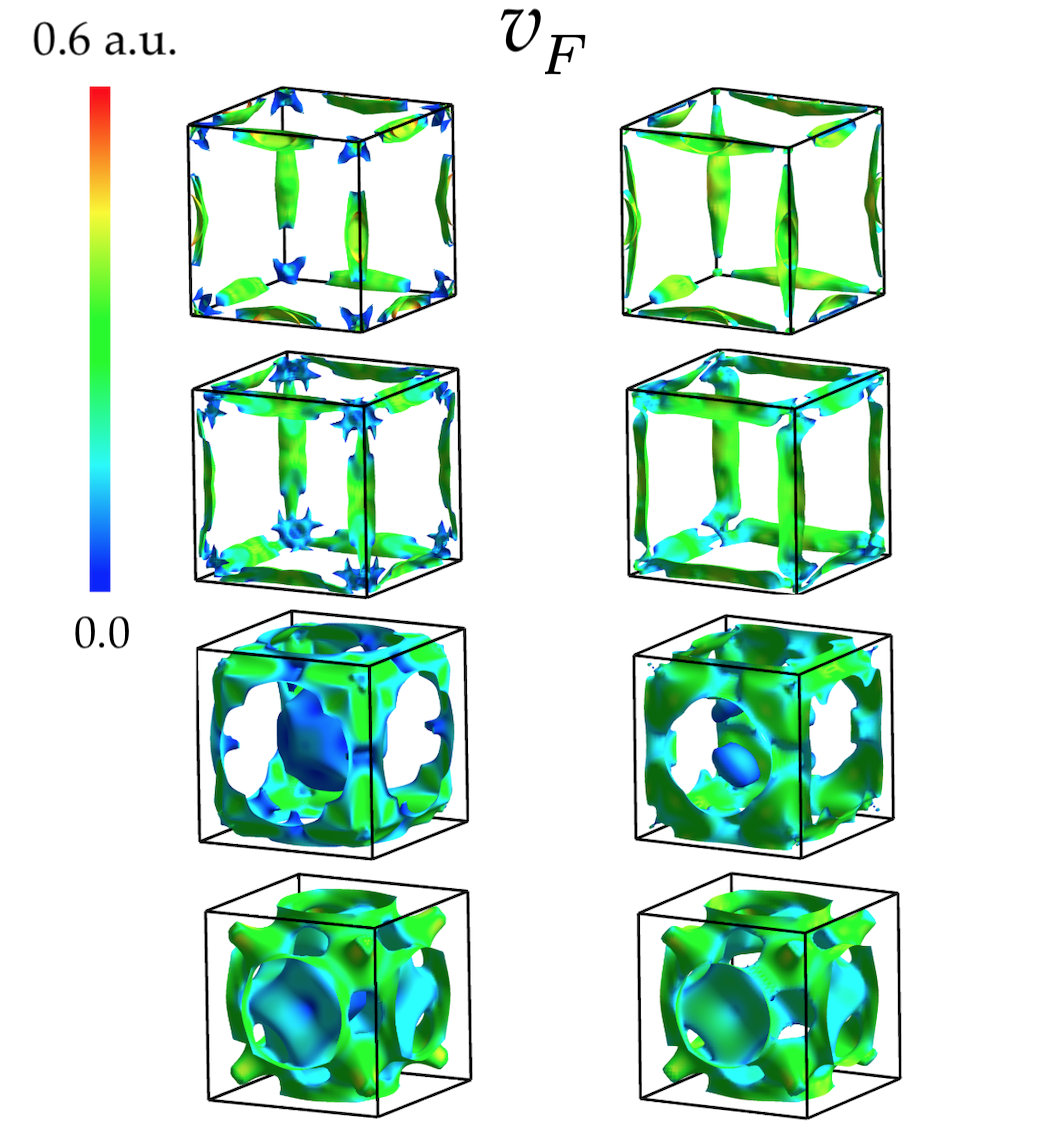}
	\caption{Fermi surface of cubic and tetragonal \ce{Nb3Sn}, broken down into single bands. Each band is decorated with Fermi velocity values. The color scale goes from zero to the maximum value of the Fermi velocity in the cubic phase.}
	\label{suppfig:FSvelocity}
\end{figure}

\newpage

\begin{figure}[ht]
    \centering
	\includegraphics[width=0.8\columnwidth]{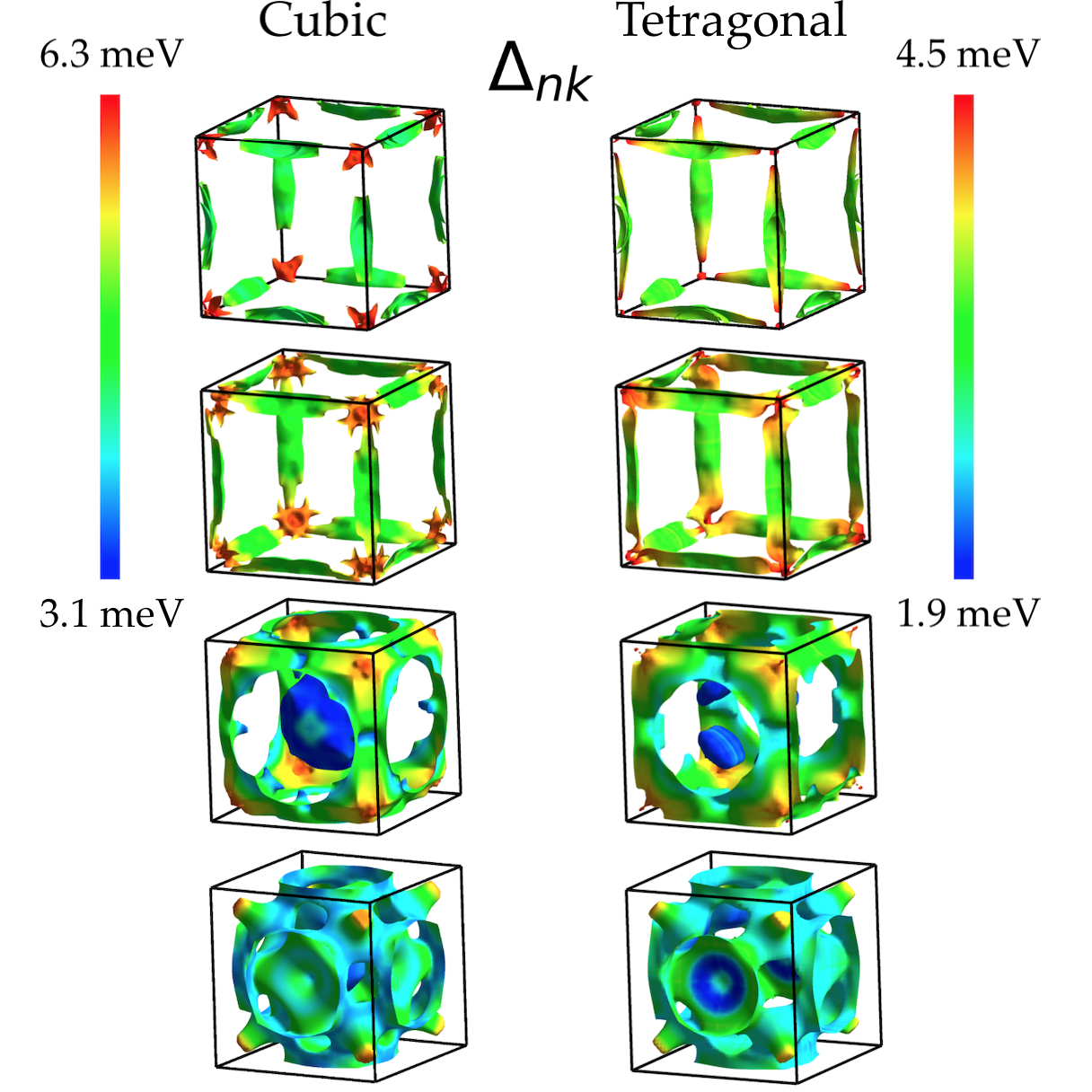}
	\caption{Fermi surface of cubic and tetragonal \ce{Nb3Sn}, broken down into single bands. Each band is decorated with anisotropic superconducting gap values at 8 K (right). The color scale goes from the minimum to the maximum value of each gap.}
	\label{suppfig:FSgap}
\end{figure}

\newpage

\begin{figure}[ht]
    \centering
	\includegraphics[width=0.8\columnwidth]{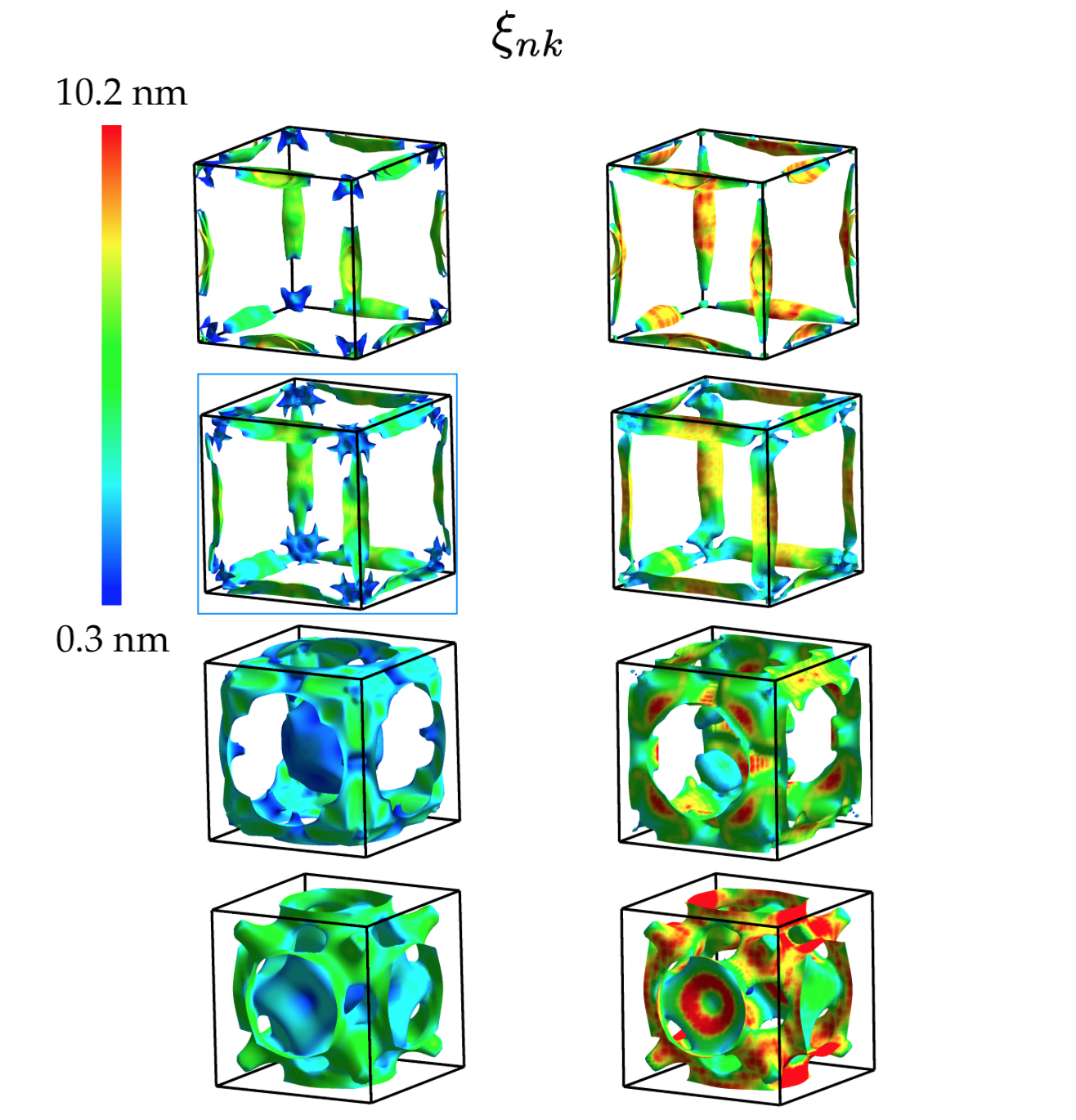}
	\caption{Fermi surface of cubic and tetragonal \ce{Nb3Sn}, broken down into single bands. Each band is decorated with coherence lenght $\xi_0(\textbf{k})$ values. The color scale goes from the minimum to the maximum value of the projection in the cubic phase.}
	\label{suppfig:FSxi}
\end{figure}

\newpage

\section{Effect of smearing on phonon dispersions}
\label{sect:Ph_smear}
\vskip10mm
\begin{figure}[ht]
    \centering
	\includegraphics[width=0.75\columnwidth]{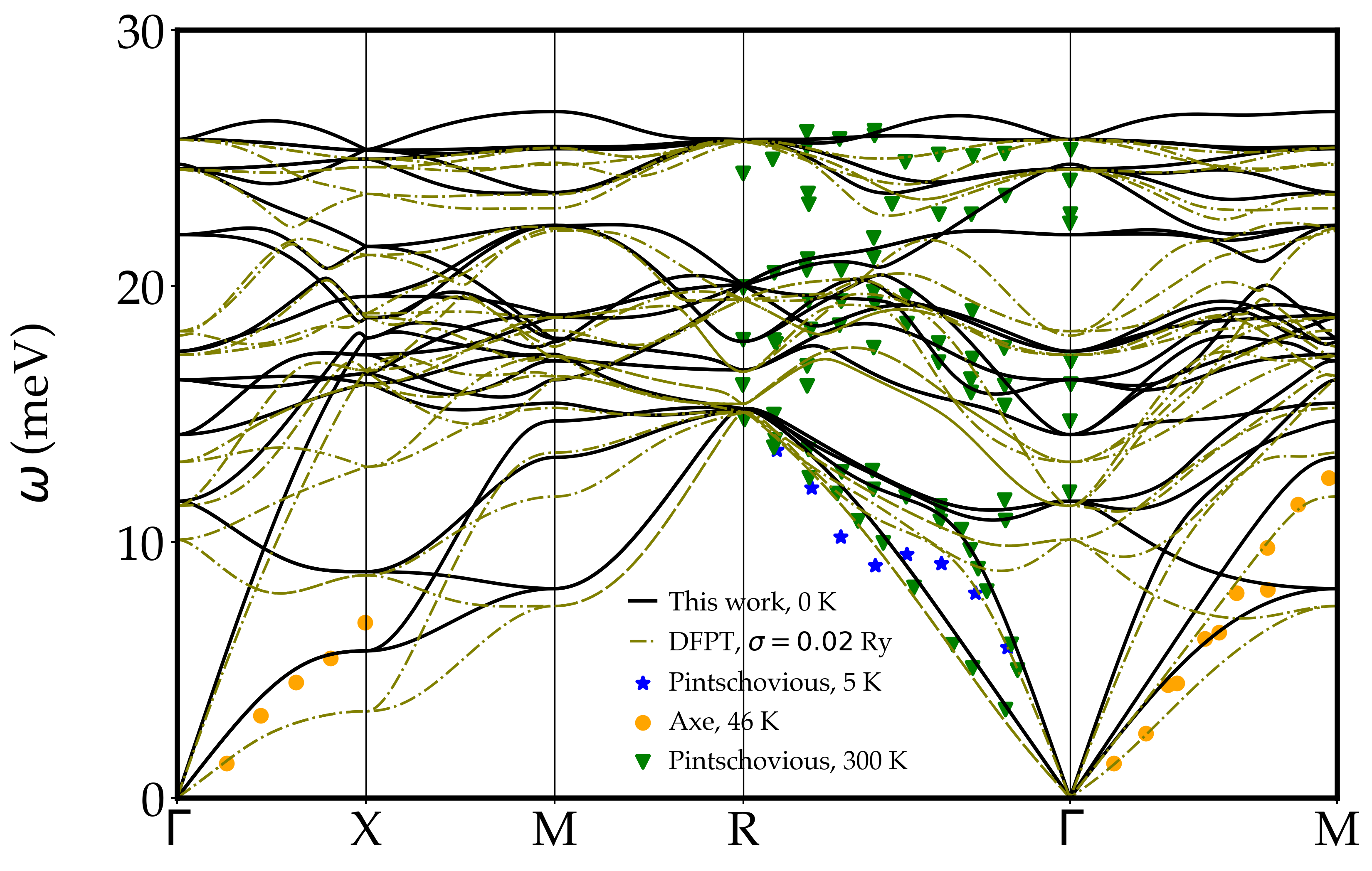}
	\caption{SSCHA phonon dispersions of cubic \ce{Nb3Sn} at $T = 0$~K compared to DFPT phonons calculated using an increased Methfessel–Paxton electronic smearing of 0.02 Ry \cite{Methfessel_PRB_1989}. Inelastic neutron scattering data from by Pintschovious \textit{et al.}~\cite{Pintschovius_phonons_nb3sn_PRB_1983, Pintschovius_phonons_nb3sn_PRL_1985} and Axe \textit{et al.}~\cite{Axe_phonons_PRB_1983} are represented by colored markers.}
	\label{suppfig:ph_smearing}
\end{figure}

\newpage

\section{Details on SSCHA optimization}
\label{sect:SSCHA_details}
In order to capture the effects of anharmonicity in phonon calculations, we employed the Stochastic Self-Consistent Harmonic Approximation (SSCHA) \cite{Errea_PRB_sscha_2014}. The SSCHA uses importance-sampling Monte Carlo, which requires the knowledge of the Born-Oppheneimer forces of an ensemble of supercell configurations. 
In the SSCHA-MLIP implementation used in this work, the forces are calculated using machine-learned interatomic potentials (MLIPs), as detailed below. The minimization of the free energy is performed with respect to the force constants $\Phi$ as implemented in the SSCHA python package \cite{Monacelli_JPCM_2021_SSCHA}. 
\\
\indent
The structure gradient is set to zero, since we do not wish to perform a structural relaxation. The average atomic positions (centroids) $R_i$ are set to coincide with the DFT equilibrium atomic positions and the initial guess for $\Phi$ is based on DFPT dynamical matrices calculated on $2\times2\times2$,  $4\times4\times4$ and $6\times6\times6$ grids. The SSCHA minimization starts with a population of 200 individuals. To ensure proper convergence, we set the ratio between the free energy gradient with respect to the auxiliary dynamical matrix and its stochastic error to be $<10^{-8}$ and the Kong-Liu ratio to be within the stochastic criterion, set at 0.5. The minimization process is then carried out using larger populations of 400 individuals. A final population of 13.000 individuals is used to extract the hessians, which is more sensible to the number of configurations. The anharmonic phonon dispersions are obtained from the positional free-energy Hessians without the fourth-order term.

\vskip10mm
\begin{figure}[ht]
    \centering
	\includegraphics[width=1.0\columnwidth]{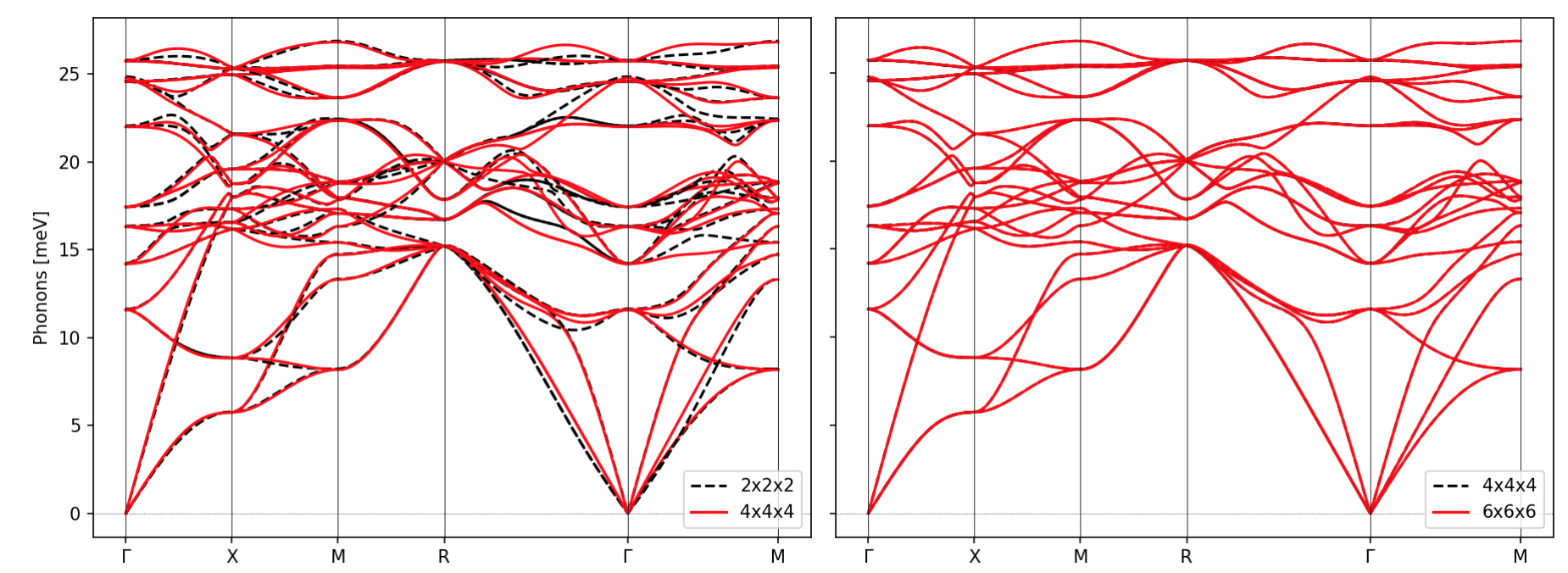}
	\caption{Comparison of the anharmonic phonon spectrum of \ce{Nb3Sn} obtained with the SSCHA, using ensembles of $2\times2\times2$,  $4\times4\times4$ and $6\times6\times6$ supercells at 0 K. }
	\label{suppfig:Supconv}
\end{figure}

\clearpage

\section{Details on MLIP training}
\label{sect:Mlip_details}
SSCHA relies on a stochastic method that requires the knowledge of the total electronic energies and forces of an ensemble of supercell configurations, typically calculated using DFT. This is the most time-consuming part of the calculation, and the current implementations based on DFT scale poorly with system size. To treat larger supercells, and ensure convergence on the results even for systems with many atoms and electrons, we employed Moment Tensor Potentials (MTPs) \cite{Novoselov_MTP_2019}. Training and evaluation of the MTPs were carried out using the MLIP package \cite{Lucrezi_MLIP_2023, Novikov_MLIP_2021, Deringer_MLIP_2019}. In order to train the potential, we generated a training set of 280 configurations and a validation set of 120 $2\times2\times2$ supercell configurations, using the SSCHA code. 
\\
\indent
We parametrized the Gaussian density matrix with the dynamical matrix of the harmonic phonon calculation and we used DFT to calculate the total energies, forces and stress tensors of the configurations. To broaden the Gaussian distribution of the generated configurations, we set the temperature of the SSCHA code to 300K. 


\newpage

\section{Solution of the Migdal-\'Eliashberg equations}
\label{sect:Solving_ME}
The self-consistent solution of the fully anisotropic Migdal-\'Eliashberg equations was performed with the EPW code \cite{Ponce_CPC_2016_EPW}, which employs maximally localized Wannier functions \cite{Giustino_PRB_2007_epw} to interpolate the electron-phonon matrix elements computed on a coarse grid on a much finer grids. Wannierization was performed using a total of 33 trial orbitals: \ce{Nb}:\{d\} for each Nb atom in the unit cell, \ce{Sn}:\{p\} for one of the two Sn atoms.
The coarse grid was $6^3$ for both cubic and tetragonal \ce{Nb3Sn}, fine \textbf{k}-meshes and \textbf{q}-meshes were $36^3/72^3$ and $18^3/52^3$ respectively for the two phases. These refined grids were subsequently utilized for integrating the electron-phonon matrix element on the Fermi surface, with an electronic smearing of 0.025 eV and a phononic smearing of 0.25 meV.

\vskip5mm
\begin{figure}[ht]
    \centering
	\includegraphics[width=0.6\columnwidth]{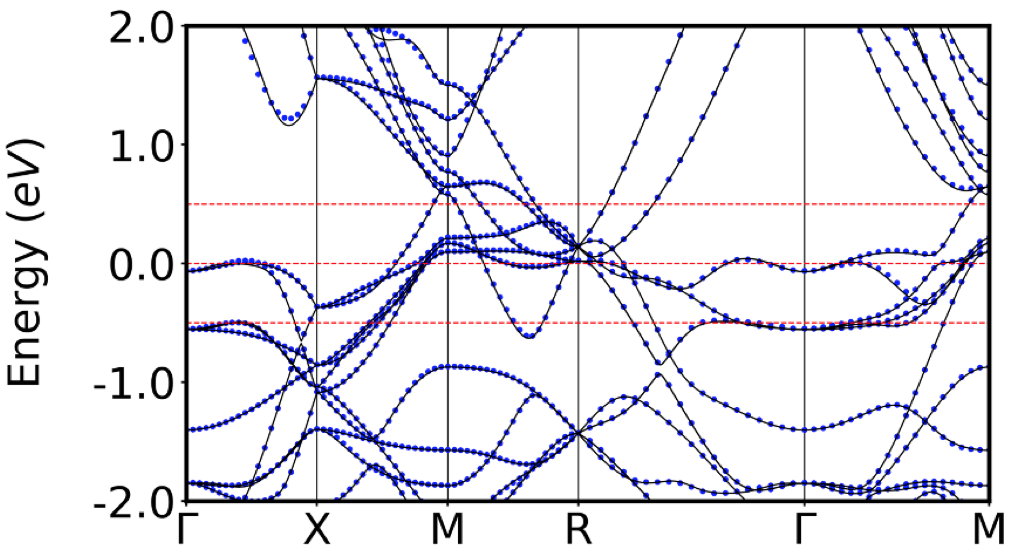}
	\caption{Comparison of the DFT electronic band structure of cubic \ce{Nb3Sn} (blue circles) and the Wannier-interpolated band structure (Solid black lines). Red dashed lines show the energy cut-off chosen to solve the Migdal-Éliashberg equations.}
	\label{suppfig:wann}
\end{figure}

We carried out the calculation with EPW's newly implemented full-bandwidth approach \cite{Lucrezi_FBW_2024}, which allows for a more accurate description of conventional superconductivity for materials with narrow electronic density of states near the Fermi level, such as is the case under investigation. Specifically, we set a Fermi surface window of width 0.3 eV. The self-consistent solution of the Migdal-Éliashberg equations was performed by fixing the chemical potential and using a Matsubara frequency cut-off of $\sim 10\hbar\,\omega_{max}$.

\indent
The extrapolation of the superconducting $T_c$ was performed solving the ME equations at different temperatures. In the anisotropic case, we then computed the weighted average of each energy distribution of the superconducting gap. Finally, the obtained data were fitted according to the following interpolation formula, built to agree with both the zero-temperature and the high-temperature limit of the superconducting gap $\Delta(T)$ in BCS superconductors:

\begin{equation}
\Delta(T) \,=\,\Delta_0\,\times\,\text{tanh}\Biggl(1.74\;\sqrt{\,\frac{T_c}{T}-1}\,\Biggr)
\label{bcs_fit}
\end{equation}
\smallskip

The numerical solution of the isotropic Migdal-\'Eliashberg beyond the $\mu^*$ approximation was performed employing in-house codes.

\begin{table}[h]
\centering
\renewcommand{\arraystretch}{1.5}
\setlength{\tabcolsep}{8pt}
\begin{tabular}{lcccc}
\hline
\hline
 & $\lambda$ & $\langle \omega^2 \rangle^{1/2}$ (meV) & $\mu^*_{\mathrm{Kresin}}$ & $\mu^*_{\mathrm{Allen\text{--}Dynes}}$ \\
\hline
Freeriks & 2.55 & 10.9 & 0.17 & 0.17 \\
Wolf     & 1.79 & 15.2 & 0.16 & 0.15 \\
Shen     & 1.56 & 13.9 & 0.09 & 0.11 \\
Geerk    & 1.50 & 13.8 & 0.08 & 0.06 \\
Rudman   & 1.75 & 14.2 & 0.13 & 0.12 \\
\hline
Average  & 1.83 & 13.6 & 0.127 & 0.122 \\
Std. dev & 0.42 & 1.6  & 0.042 & 0.044 \\
\hline
This work (cubic) & 2.08 & 17.4 & - & 0.22 \\
\hline
This work (tetragonal) & 1.52 & 16.6 & - & 0.19 \\
\hline
\end{tabular}
\caption{Electron-phonon coupling constants $\lambda$, averaged phonon frequencies $\langle \omega^2 \rangle^{1/2}$, and corresponding Coulomb pseudopotentials $\mu^*$ extracted using the Kresin and Allen–Dynes formalisms in Refs. \cite{Shen_nb3sn_a2f_PRL_1972, Rudman_nb3sn_a2f_1984, Freericks_nb3sn_tunnelling_PRB_2002, Kieselmann_nb3sn_tunnelling_1982, Wolf_tunnelling_a2F_2011, Geerk_tunnelling_a15_1985}. The row labeled “This work” refers to our results for cubic and tetragonal \ce{Nb3Sn} derived from KO Coulomb interaction.}
\label{tab:mu_star_list}
\end{table}

\vskip5mm
\begin{table}[h]
\centering
\renewcommand{\arraystretch}{1.4}
\setlength{\tabcolsep}{8pt}
\begin{tabular}{lccccc}
\hline
\hline
Phase & $\mu^*_{\,\mathrm{RPA}}$ & $T_c^{\,\mathrm{RPA}}$ (K) & $\mu^*_{\,\mathrm{KO}}$ & $T_c^{\,\mathrm{KO}}$ (K) & $T_c^{\,\mathrm{exp}}$\\
\hline
Cubic      & 0.17 & 24.7 & 0.22 & 21.8 & 18\\
Tetragonal & 0.15 & 18.4 & 0.19 & 16.2 & 17\\
\hline
\hline
\end{tabular}
\caption{Screened Coulomb pseudopotentials ($\mu^*$) obtained from RPA and Kukkonen--Overhauser (KO) calculations \cite{Pellegrini_Nb_PRB_2023}, and corresponding critical temperatures $T_c$ from the anisotropic Migdal--Eliashberg equations. Experimental $T_c$ values are reported for comparison.}
\label{tab:tc_table}
\end{table}

\newpage

\vskip5mm
\begin{figure}[h]
    \centering
	\includegraphics[width=1.0\columnwidth]{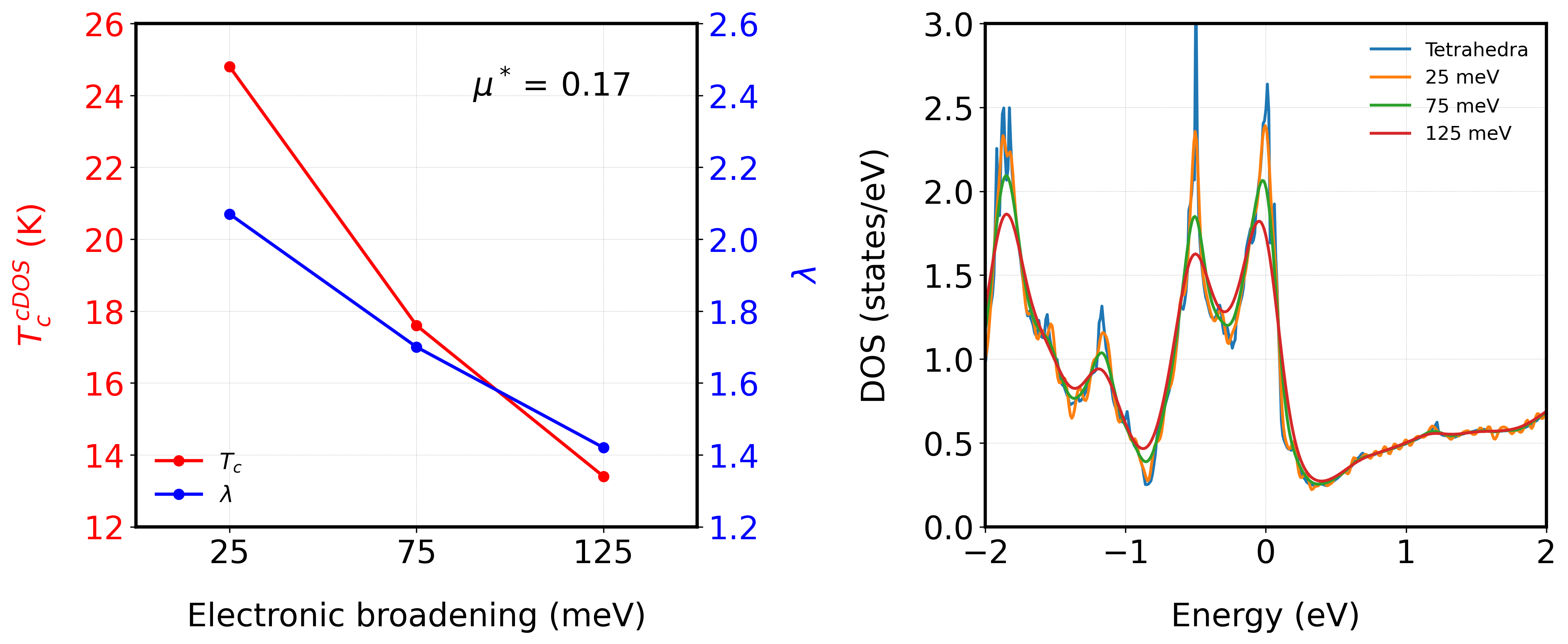}
	\caption{(a) Convergence of electron-phonon coupling constant $\lambda$ and $T_c$ with respect to electronic smearing from c-DOS solution of the isotropic ME equations. (b) Smoothening of the DOS profile of cubic \ce{Nb3SN} due to electronic smearing.}
	\label{suppfig:wann}
\end{figure}

\newpage

\section{DFPT calculations - convergence tests}
\label{sect:conv_test}

\begin{figure}[ht]
    \centering
	\includegraphics[width=0.8\columnwidth]{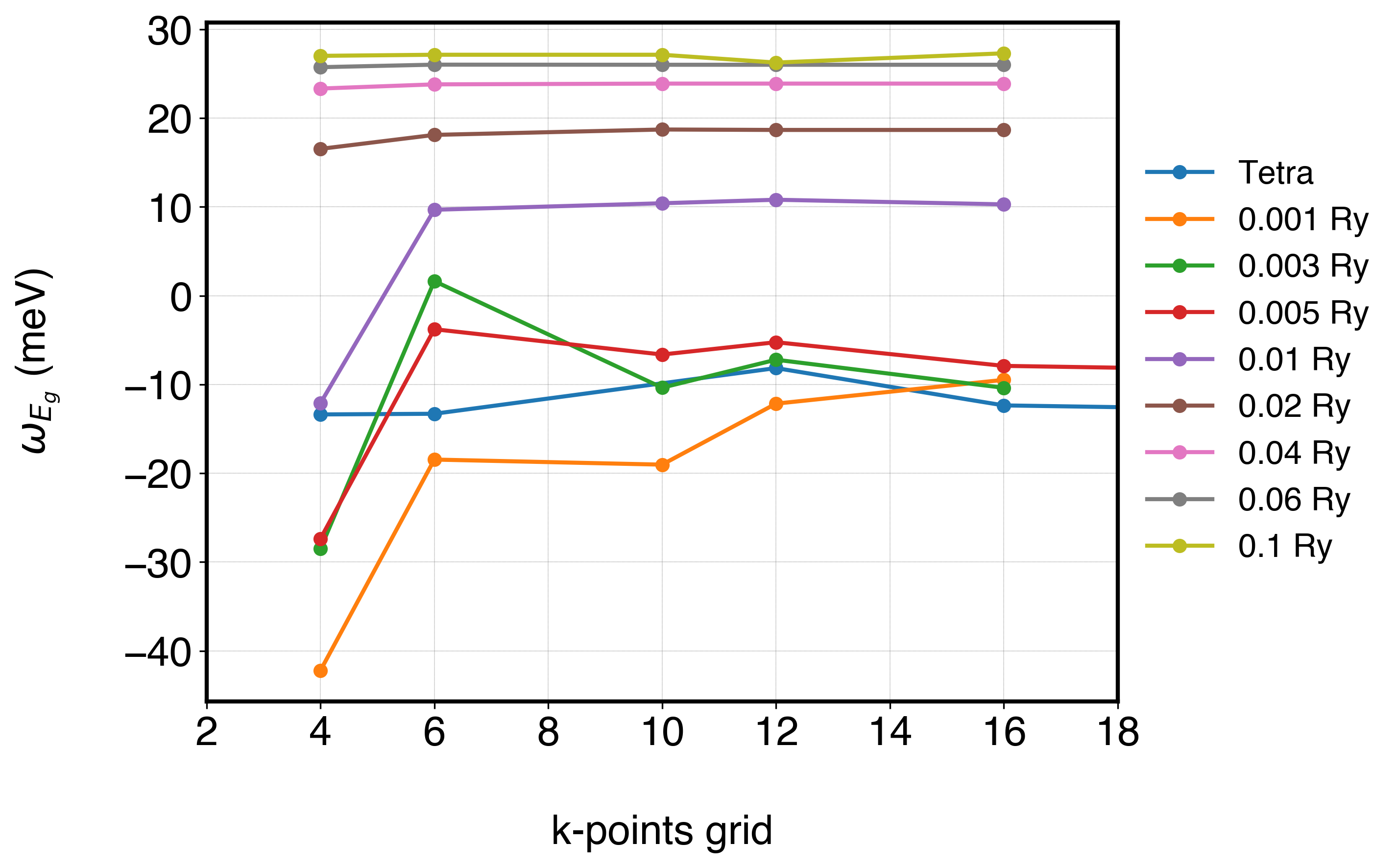}
	\caption{Convergence of the zone-center $\Gamma_{12}^+$ phonon mode of cubic \ce{Nb3Sn} with respect to uniform \textbf{k}-point grids in reciprocal space, for different values of the electronic smearing. Blue points are calculated using the optimized tetrahedron method.}
	\label{suppfig:kconv}
\end{figure}

\clearpage
\bibliographystyle{apsrev4-1}
\bibliography{library}